\documentclass[12pt]{article}

\usepackage{amsmath,amssymb}
\usepackage[scr=boondox]{mathalpha}
\usepackage[T1]{fontenc}
\usepackage[utf8]{inputenc}
\usepackage{textcomp} 
\usepackage{parskip}
\usepackage{xcolor}
\usepackage[margin=1in]{geometry}
\usepackage{graphicx}
\usepackage{setspace,lscape} 
\usepackage{caption,subcaption,multirow} 
\usepackage[bottom]{footmisc}
\usepackage{hyperref}
\usepackage{enumitem} 
\usepackage{standalone}

\usepackage{natbib}
\renewcommand{\[}{\begin{equation}} 
\renewcommand{\]}{\end{equation}}

\title{Toward an Attentional Model of Time Discounting}
\author{Zijian Zark Wang\thanks{Warwick Business School, University of Warwick. Email: \texttt{zijianwang1996@gmail.com}. I am grateful to Daniel Read for his advice through this project. I thank the comments of Pedro Bordalo, Zvi Safra, Andrea Isoni, Tim Mullet, Graham Loomes, Giovanni Burro and seminar participants at the SABE-IAREP conference, EIF CogAI, China BEEF, Warwick Business School, and Zhejiang University. All errors are my own.} 
}
\date{\today}

\begin{document} 
\maketitle 
\begin{abstract} 
This paper proposes a new theory of time discounting. We assume that when decision makers evaluate a sequence of future rewards, they allocate more attention to large rewards and less attention to small rewards. Moreover, when attention is distributed over a longer time horizon, each period receives less attention. The reductions in attention lead to greater discounting of reward values, and the resulting discount factors follow a distribution similar to the multinomial logit function. We characterize such discount factors using two approaches: an axiomatic analysis based on the optimal discounting framework, and a neuro-computational mechanism based on a simple spiking neural network. The theory explains a wide range of anomalies, including the hidden-zero effect, S-shaped value function, intertemporal correlation aversion, concentration bias, and inconsistent planning. Also, it identifies new determinants for several well-documented phenomena, such as the common difference effect and risk aversion over time lotteries. 
\end{abstract}

\section{Introduction}\label{sec:introduction}

Suppose a decision maker (DM) faces a sequence of rewards delivered at different times. In economics, the value of such a reward sequence is typically modeled using the additive discounted utility framework \citep[see][]{cohen2020measuring}. Classical models in this framework, including exponential and hyperbolic discounting, assume that how a future reward is discounted depends only on its timing. However, a growing body of evidence suggests that the valuation of a reward is often shaped by the amount and timing of other rewards in the same sequence \citep{loewenstein1993preferences,urminsky2011scope,read2012tradeoffs,scholten2016cumulative,scholten2024unified}. In this paper, we propose that attention plays a key role in such contextual influences.

Existing research on attention in psychology has largely focused on visual attention. Three well-established findings from this literature motivate our research. First, attention is driven by reward values. In visual search tasks, visual cues previously associated with high reward value automatically capture attention. As the reward is larger, it captures attention more strongly \citep{hickey2010reward,anderson2011value,chelazzi2013rewards}. Second, attention is limited. When only one visual cue appears in view, people can easily focus on it and discriminate its features. In contrast, when many closely spaced visual cues appear simultaneously, attention to each cue is reduced, and its features are perceived less precisely \citep{he1996attentional,intriligator2001spatial,whitney2011visual}. Third, attention can amplify values. When people choose between two rewards, directing their attention to one option could make it valued higher \citep{armel2008biasing,gwinn2019spillover,smith2019gaze,pleskac2023attention}. We hypothesize that these mechanisms of visual attention also operate when people evaluate sequences of rewards.

To illustrate, consider a reward sequence in which you receive £30 today and £20 in one month. We then change this sequence in two ways. First, imagine that the £20 reward is increased to £3,020. When evaluating the new sequence ``receive £30 today and £3,020 in a month'', the larger future reward may capture most of your attention. In this situation, the £30 reward may receive shallower processing, leading it to be undervalued or underweighted relative to how it is valued in the original sequence.

Second, imagine that you have to process more information in the sequence. For instance, the original sequence is expanded to ``receive £30 today and £20 in a month, and £10 in two months, and £15 in three months, and £25 in four months''. Evaluating this extended sequence requires processing all the newly added rewards. As the number of rewards increases, each individual reward may receive less attention due to limited attention capacity. Although each reward is clearly visible, its depth of processing may decline, leading it to be undervalued.


In both cases, reduced attention to the £30 reward may result in greater discounting of its value (note the value-amplifying role of attention). We develop a new model of time discounting based on this intuition, and show that the model can explain a wide range of decision anomalies. 

We assume that in intertemporal choice, the DM evaluates each sequence separately. Let \(s_{0\to T}=[s_0,s_1,…,s_T]\) denote a sequence of rewards delivered from period 0 to period \(T\), where \(s_t\) represents the reward in period \(t\). In the additive discounted utility framework, the sequence's value can be represented by \(U(s_{0\to T})=\sum_{t=0}^T w_t u(s_t)\), where \(u(s_t)\) is the instantaneous utility, and \(w_t\) is the decision weight assigned to it. Importantly, we recommend using a simple form to specify \(w_t\):
\begin{equation*}
    w_t = \frac{d_t \,e^{u(s_t)/\lambda}}{\sum_{i=0}^T d_i \,e^{u(s_i)/\lambda}}
\end{equation*}
where \(\lambda>0\), and \(d_t\) represents the DM's baseline patience or the discount factor of period \(t\) before she acquires information about the sequence. Through most of this paper, we assume that \(d_t\) is the exponential discount factor, so that decision anomalies arise only from the attention mechanisms that covert $d_t$ into $w_t$. 

Our model has three desired features. First, \(w_t\) captures the ideas we just illustrated in the two cases. It increases with reward \(s_t\) but decreases with other rewards, indicating a tendency to focus on larger rewards (as in the first case). Also, the decision weights sum to a fixed amount, indicating that the DM's attention capacity is limited, and that when attention is divided over more periods, each period receives less attention (as in the second case).




 Second, we introduce only one additional parameter \(\lambda\) to the standard additive discounted utility framework. This makes our model not only capable of explaining a range of phenomena, but also highly parsimonious. Meanwhile, the parameter \(\lambda\), as discussed later, is also meaningful. It decreases with the DM's learning rate and increases with the cognitive cost of reallocating attention. 

Third, \(w_t\) follows a  the multinomial logit (also called softmax) function, which is widely used to model attention in various research fields. In discrete choice analysis, the softmax function can represent the DM's choice strategy under attention constraint \citep{matvejka2015rational}; in memory research, \citet{bordalo2020memory} use this function to model attention weights for recalled experiences. In machine learning, the softmax function is widely applied to represent attention weights in image processing and language processing \citep{bahdanau2015iclr-neural,xu2015show,vaswani2017attention,dosovitskiy2021image}. Therefore, our model has the potential to be integrated with other models to form a unified framework for analyzing the role of attention in decision-making.

\subsection{Two Characterization Approaches}\label{two-characterization-approaches}

We characterize our model using two approaches. The first approach is an axiomatic analysis based on the optimal discounting framework \citep{noor2022optimal,noor2024constrained}. The optimal discounting model assumes that, in evaluating a reward sequence, the DM adjusts decision weights to maximize the net value she perceives. Adjusting such weights incurs a cognitive cost, so the net value she perceives equals the total sequence value minus the cognitive cost. We show that within this framework, our proposed weighting function is the only function that satisfies the standard preference axioms as well as two additional axioms: Sequential Outcome-Betweenness, and Bracketing Invariance.

According to Sequential Outcome-Betweenness, adding a new reward to the end of a sequence does not always increase the sequence value. If the new reward is tiny, the sequence value may decrease. This axiom is supported by experimental evidence \citep{scholten2014better, jiang2017better}. According to Bracketing Invariance, when valuing a sequence, the decision weight assigned to a reward is invariant to whether the remaining part of the sequence is viewed as a combination of individual rewards or as a single bracketed unit. This axiom ensures that for any particular sequence, the decision weights can be uniquely determined. 



The second approach is based on spiking neural networks (SNNs). SNNs are neural models that mimic the information-processing mechanisms of biological neural systems. Such models exhibit high biological plausibility and have attracted considerable interest from machine learning researchers \citep{maass1997networks,gerstner2002spiking,tavanaei2019deep}. In neuroscience, the connection between two neurons is called a synapse, and the strength of a synapse is called the synaptic weight. We design a simple two-layer SNN in which the synaptic weights between the layers follow two basic neuro-computational principles. The first is Hebbian learning \citep{hebb1949}: a synapse is strengthened when the two neurons it connects are co-activated. The second is homeostatic scaling \citep{turrigiano2000hebb,turrigiano2008self}, which stabilizes neuronal activity by scaling all synaptic weights up or down by a common factor. We show that these two principles co-drive the expected synaptic weights to evolve approximately according to our proposed weighting function. To the best of our knowledge, this paper is the first to use SNN to develop a behavioral economic theory.  


\subsection{Behavioral Implications and Related Literature}\label{applications-and-related-literature}

Our model explains a range of decision anomalies, including S-shaped value function, inconsistent planning \citep{laibson1997golden}, the hidden zero effect \citep{magen2008hidden}, intertemporal correlation aversion \citep{andersen2018multiattribute}, and concentration bias \citep{dertwinkel2022concentration}. The parameter \(\lambda\) plays a key role in generating these anomalies. For example, the emergence of S-shaped value function and concentration bias requires a low \(\lambda\), while intertemporal correlation aversion requires a high \(\lambda\). In general, this suggests that the strength and frequency of many anomalies in intertemporal choice may be moderated by the DM's rate or the cost of learning. 


The model makes novel predictions about some well-documented behavioral effects, such as the common difference effect \citep{loewenstein1992anomalies} and risk aversion over time lotteries \citep{onay2007intertemporal, dejarnette2020time}. It predicts that in comparison between a small sooner reward (SS) and a large later reward (LL), if the difference between reward delays is substantially larger than the difference between reward utilities, the common difference effect may be reversed. That is, adding a common delay to both SS and LL may lead people to exhibit less patience. Besides, it predicts that people are risk averse over time lotteries only when the lottery reward is large enough.


In recent years, there has been growing interest in the role of attention in economic decisions, including intertemporal choice. Our model is related to this literature. The closest theories to ours are salience theory \citep{bordalo2012salience} and focus-weighted utility theory \citep{kHoszegi2013model}. Their assumptions likewise lead to greater decision weights on larger rewards. However, they do not impose a constraint on the total weight, which yields different predictions from ours. \citet{steiner2017rational} introduce rational inattention to explain anomalies in dynamic intertemporal decision making. They assume that the DM acquires limited information at each decision step. We additionally consider how the DM would allocate attention to different pieces of information. \citet{noor2022optimal} propose that paying attention to the future triggers a cognitive cost, but they do not specify how the value of the present reward changes when a future reward captures attention. In Section \ref{relation-to-other-intertemporal-choice-models}, we provide a more detailed comparison with some related theories. 

On the empirical side, \citet{fisher2021intertemporal} show that attention can causally influence intertemporal choice using eye-tracking experiments. Also, several other studies provide correlational evidence \citep{franco2016now,amasino2019amount,keidel2025attentional,chen2026attentional,kang2026relationship}. These findings suggest that attention-based explanations may offer a promising and testable account of intertemporal choice, although most studies involve only single-period rewards.


The rest of this paper is organized as follows. Section \ref{sec:model_setting} introduces the model setting. Section \ref{sec:characterization} describes the two approaches used to characterize the model. Section \ref{sec:implications} explains the behavioral implications of the model. Section \ref{sec:model_discuss} discusses issues related to the use of the model and compare the model with other studies. Section \ref{sec:conclusion} concludes.

\section{Model Setting} \label{sec:model_setting}

Assume time is discrete. Let \(s_{0\to T}\equiv[s_0,s_1,...,s_T]\) denote a reward sequence that starts delivering rewards at period 0 and ends at period \(T\), and \(s_t\) denote the reward delivered at period \(t\) (\(s_t\) can be zero). Through this paper, we only consider non-negative rewards and finite length of sequence, i.e.~\(s_t \in \mathbb{R}_{\geq 0}\) and \(1\leq T<\infty\). The DM's choice set is constituted by a set of reward sequences that start from period 0 and may end at different periods. The value of \(s_{0\rightarrow T}\) is defined as \(U(s_{0\rightarrow T})\equiv \sum_{t=0}^T w_{t}u(s_t)\), where \(u(s_t)\) is the instantaneous utility of \(s_t\), and \(w_t\) is the decision weight (\(0<w_t<1\)) assigned to \(s_t\). The function \(u(.)\) is strictly increasing and unbounded above; for convenience, we set \(u(0)=0\).

The determination of \(w_t\) is central to this paper. We assume that when evaluating a reward sequence, the DM divides her attention to each period in the sequence, in order to acquire the related utility information. The more attention she pays to a reward, the greater decision weight is assigned to that. In intuition, to evaluate reward \(s_t\), she may imagine how much pleasure she would feel on the occasion when she receives \(s_t\) \citep{macinnis1987role}. As more attention is paid to that specific occasion, her imagination becomes more vivid and hence the perceived pleasure (or the value of \(s_t\)) \(w_t u(s_t)\)  increases. We assume that if the DM is totally focused on that specific occasion, the value of \(s_t\) will be equal to its instantaneous utility. After going through all periods, the DM aggregates the perceived reward values to construct a representation of the total sequence value.


The division of attention is subject to all rewards in the sequence. Specifically, we propose two attention allocation principles. First, the DM pays more attention to large rewards and less attention to small rewards. For example, consider a reward sequence ``£5 today and £200 in 1 month''. When valuing this sequence, the DM may attend more to £200 and less to £5. Second, when the DM has to value more rewards in a sequence, the attention allocated to each reward declines. To illustrate, suppose we extend the former sequence to ``£5 today and £200 in 1 month, and £85 in 2 months, and £10 in 3 months''. To value the new sequence, the DM needs to take into account more rewards. She may therefore pay less attention to £5 and £200 than in the former case, and the occasion imagined for receiving each reward may be less vivid. In Section \ref{sec:implications}, we show that these two principles can account for a wide range of anomalies related to intertemporal choice. 

Specifically, we assume that the decision weight \(w_t\) follows a softmax function. We define any weight in this style as an \emph{attention-modulated discount} (AMD) factor.

\noindent \textbf{Definition 1:} \emph{Let \(\mathcal{W}\equiv[w_0,...,w_T]\) denote the decision weights for all rewards in sequence \(s_{0\rightarrow T}\). \(\mathcal{W}\) is called attention-modulated discount factors if for any \(t\in\{0,1,…,T\}\),
\[\label{eq:def_amd}
w_t = \frac{d_t\,e^{u(s_t)/\lambda}}{\sum_{j=0}^T d_j \,e^{u(s_j)/\lambda}} 
\]
where \(0< d_t \le 1\) and \(\lambda>0\).}


How Definition 1 reflects the role of attention can be explained in three points. First, the decision weights are normalized so that their sum is constant. This indicates that attention is limited. In this paper, this normalizing constraint has two implications. On the one hand, increasing the decision weight of one reward reduces the decision weights of other rewards in the sequence, implying that focusing on one item could make the DM less sensitive to others. On the other hand, when there are more periods in the sequence, the DM needs to divide attention across a longer horizon, which may reduce the attention to, or decision weight of, each individual reward.

Second, \(w_t\) is strictly increasing in \(s_t\), indicating a tendency to pay more attention to larger rewards. This is consistent with empirical findings in many decision domains. For example, value-driven attentional capture \citep{hickey2010reward, anderson2011value, chelazzi2013rewards} shows that visual stimuli associated with large rewards naturally capture attention. Meanwhile, the ostrich effect \citep{galai2006ostrich, karlsson2009ostrich} shows that people attend more to good news and less to bad news.




Third, \(w_t\) is ``anchored'' in a factor \(d_t\), which indicates that attention is accompanied by cognitive costs. The factor \(d_t\) can represent the DM's baseline patience; thus, we term it the default discount factor. The distribution of \(d_t\) determines the initial decision weight that the DM would assign to each reward without knowing its realization. The DM reallocates attention as she learns the realization of each reward. This attention reallocation process is cognitively costly. The deviation of \(w_t\) from \(d_t\) is mediated by a parameter \(\lambda\), which could represent the cost of attention reallocation. The greater \(\lambda\) is, the closer \(w_t\) is to \(d_t\).


\section{Characterization} \label{sec:characterization}

Some researchers may confuse Equation (\ref{eq:def_amd}) with the multinomial logit choice model. We note here that \(w_t\) is not the choice probability of an option; rather, it is the decision weight used to compute the option value. Thus, the characterization approaches proposed for the multinomial logit choice model \citep[e.g.][]{matvejka2015rational,caplin2022rationally,cerreia2023multinomial} are not directly applicable to our model. We present two alternative approaches to characterize the AMD factors. The first is to add extra axioms to the optimal discounting model \citep{noor2022optimal,noor2024constrained}. The second is to approximate via the learning dynamics of a simple spiking neural network.

\subsection{Optimal Discounting}\label{optimal-discounting}

The optimal discounting model assumes that, in evaluating a reward sequence, the DM subjectively adjusts the decision weights so as to maximize the perceived total value of the sequence. This weight adjustment process triggers a cognitive cost. The DM needs to trade off the increase in perceived sequence value against the cognitive cost.  

There are two attention-based explanations for why the DM adjusts the decision weights to maximize the perceived value. Both explanations reflect a psychological view that information carries not only instrumental value (helping identify the optimal action) but also hedonic value (yielding pleasure) and cognitive value (satisfying curiosity) \citep{gottlieb2010attention,gottlieb2012attention,sharot2020people}. The first explanation is that the DM seeks to acquire as much pleasant information as possible. The second is that, driven by curiosity, the DM seeks to acquire as much surprising information as possible; if, before observing the sequence, the DM expects zero value, then the total sequence value can serve as a proxy for the degree of surprise.\footnote{Many decision theories have been built on these two explanations, such as \cite{ely2015suspense} and \cite{golman2016information}. In addition, theories based on the first explanation also include \cite{loewenstein1987anticipation} and \cite{brunnermeier2017optimal}, while those based on the second also include \cite{itti2009bayesian} and \cite{friston2017active}.} Under these explanations, information about larger rewards should attract more of her attention, as it can bring greater pleasure and contribute more to the total surprise.

\citet{noor2022optimal,noor2024constrained} build a framework for the optimal discounting problem. We adopt a variant of their original model and specify it in Definition 2.

\noindent \textbf{Definition 2}: \emph{Given a reward sequence \(s_{0\to T}=[s_0,...,s_T]\), the following optimization problem is called an optimal discounting problem for \(s_{0\to T}\):
\[\label{eq:def_optimal_discount}
\begin{aligned} 
&\max_{\mathcal{W}}\;&&\sum_{t=0}^T w_tu(s_t) - C(\mathcal{W}) \\ 
&s.t.\; &&\sum_{t=0}^Tw_t \leq M \\
&&& w_t \geq 0 \text{ for all } t\in \{0,1,...,T\} 
\end{aligned} 
\]
where \(M>0\), \(u(s_t)<\infty\). \(C(\mathcal{W})\) is the cognitive cost function and is constituted by time-separable costs, i.e. \(C(\mathcal{W})=\sum_{t=0}^Tf_t(w_t)\). For all \(w_t\in(0,1)\), \(f_t(w_t)\) is differentiable, \(f'_t(w_t)\) is continuous and strictly increasing, and \(\lim_{w_t\rightarrow 0} f'_t(w_t)=-\infty\).
}

In Definition 2, the DM's objective is to maximize the total sequence value minus the cognitive cost. The DM has a limited capacity of attention, which is captured by $M$. In the solution to Equation (\ref{eq:def_optimal_discount}), \(w_t\) increases with \(s_t\), which can capture the DM's tendency to pay more attention to larger rewards. 

There are three differences between Definition 2 and the original model of \cite{noor2022optimal,noor2024constrained}. First, in our setting, shifting attention to future rewards may reduce the attention to the current reward, while this would never happen in \citet{noor2022optimal,noor2024constrained}. Second, the original model assumes \(f'_t(w_t)\) must be continuous at 0 and \(w_t\) must be no larger than 1. We relax these assumptions since neither \(w_t=0\) nor \(w_t\geq1\) is included in our solutions. Third, the original model assumes that \(f'_t(w_t)\) is left-continuous in \([0,1]\), and there exist \(\underline{w},\bar{w}\in[0,1]\) such that \(f'_t(w_t)=0\) when \(w_t\leq\underline{w}\), \(f'_t(w_t)=\infty\) when \(w_t\geq\bar{w}\), and \(f'_t(w_t)\) is strictly increasing when \(w_t \in [\underline{w},\bar{w}]\). We simplify this assumption by setting \(f'_t(w_t)\) to be continuous and strictly increasing in \((0,1)\), and similarly, we assume \(f'_t(w_t)\) can approach infinity near at least one border of \([0,1]\).\footnote{For convenience, we set \(\lim_{w_t\rightarrow 0} f'_t(w_t)=-\infty\), but it will be fine to assume it can approach positive infinity near the other border.}






Let \(\succsim\) denote the preference relation between two reward sequences.\footnote{If \(a \succsim b\) and \(b\succsim a\), we say \(a\sim b\). If \(a \succsim b\) does not hold, we say \(b\succ a\). \(\succsim\) can also characterize the preference relation between single rewards as any single reward can be viewed as a one-period sequence.} Besides, for any reward sequence \(s_{0\to T}=[s_0,...,s_T]\), we define \(s_{0\to t}=[s_0,...,s_t]\) as a sub-sequence of it, where \(1\leq t\leq T\). Unless otherwise specified, every sub-sequence starts from period 0. 

We first introduce two axioms for \(\succsim\):

\noindent \textbf{Axiom 1:} \emph{\(\succsim\) has the following properties:
\begin{enumerate} \def\labelenumi{(\alph{enumi})} 
\item (Complete Order) \(\succsim\) is complete and transitive. 
\item (Continuity) For any reward sequences \(s,s'\) and reward \(c\in \mathbb{R}_{\geq 0}\), the sets \(\{\alpha \in(0,1) | \alpha\cdot s + (1-\alpha)\cdot c \succsim s'\}\) and \(\{\alpha \in(0,1) | s' \succsim \alpha\cdot s + (1-\alpha)\cdot c \}\) are closed. 
\item (Independence) For any reward sequences \(s,s'\) and reward \(c\in \mathbb{R}_{\geq 0}\), \(s \succsim s'\) implies that for any \(\alpha \in (0,1)\), \(\alpha\cdot s + (1-\alpha)\cdot c \succsim \alpha \cdot s' + (1-\alpha) \cdot c\). 
\item (Reduction of Compound Alternatives) For any reward sequences \(s,s',q\) and rewards \(c_1,c_2\in \mathbb{R}_{\geq 0}\), if there exist \(\alpha, \beta \in (0,1)\) such that \(s \sim \alpha \cdot q + (1-\alpha) \cdot c_1\), then \(s' \sim \beta \cdot q + (1-\beta)\cdot c_2\) implies \(s' \sim \beta\alpha\cdot q+\beta(1-\alpha)\cdot c_1 + (1-\beta)\cdot c_2\). 
\end{enumerate}
}

\noindent \textbf{Axiom 2:} \emph{For any \(s_{0\to T}\) and any \(\alpha_1,\alpha_2 \in (0,1)\), there exists \(c\in \mathbb{R}_{\geq 0}\) such that \(\alpha_1 \cdot s_{0\to T-1}+\alpha_2\cdot s_T \sim c\).}

These two axioms are almost standard in decision theories. The assumption of complete order implies that preferences between reward sequences can be characterized by a utility function. Continuity and independence ensure that in a stochastic setting where the DM can receive one reward sequence under some states and receive a single reward under other states, her preference can be directly characterized by the expected utility theorem. Reduction of compound alternatives ensures that the valuation of a specific sequence is constant over states. Axiom 2 is an extension of the constant-equivalence assumption in \citet{bleichrodt2008koopmans}. It implies that there always exists a constant that can represent the value of a convex combination of sub-sequence \(s_{0\to T-1}\) and reward \(s_T\).

Suppose \([w_0,...,w_T]\) is a solution to the optimal discounting problem. In the optimal discounting framework, the DM's preference for \(s_{0\to T}\) can be characterized by the preference for \(w_0\cdot s_0+w_1\cdot s_1 +...+w_T\cdot s_T\). We specify this assumption in Definition 3.

\noindent \textbf{Definition 3:} \emph{Given reward sequence \(s_{0\to T}=[s_0,...,s_T]\) and \(s'_{0\to T'}=[s'_0,...,s'_{T'}]\), the preference relation \(\succsim\) admits an optimal discounting representation if
\begin{equation*}
    s_{0\to T} \succsim s'_{0\to T'}\quad \Longleftrightarrow \quad \sum_{t=0}^T w_t\cdot s_t \succsim \sum_{t=0}^{T'} w'_t \cdot s'_t 
\end{equation*}
where \(\{w_t\}_{t=0}^T\) and \(\{w'_t\}^{T'}_{t=0}\) are solutions to the optimal discounting problems for \(s_{0\to T}\) and \(s'_{0\to T'}\) respectively.}\footnote{\citet{noor2022optimal} refer to the term ``optimal discounting representation'' as Costly Empathy representation.}

Furthermore, if Definition 3 is satisfied and both \(\{w_t\}_{t=0}^T\) and \(\{w'_t\}^{T'}_{t=0}\) take the AMD form, we say \(\succsim\) admits an AMD representation. Now we specify two additional behavioral axioms that are key to characterizing AMD.

\noindent \textbf{Axiom 3:} (Sequential Outcome-Betweenness) \emph{For any \(s_{0\to T}\), there exists \(\alpha\in(0,1)\) such that \(s_{0\to T} \sim \alpha\cdot s_{0\to T-1}+(1-\alpha) \cdot s_T\).}

\noindent \textbf{Axiom 4:} (Bracketing Invariance) \emph{Suppose \(T\geq 2\). For any \(s_{0\to T}\), if there exist \(\alpha_1,\alpha_2,\beta_0,\beta_1,\beta_2\in(0,1)\) such that \(s_{0\to T}\sim \alpha_1 \cdot s_{0\to T-1} + \alpha_2 \cdot s_{T}\) and \(s_{0\to T}\sim \beta_0 \cdot s_{0\to T-2}+\beta_1 \cdot s_{T-1}+\beta_2 \cdot s_{T}\), then we must have \(\alpha_2 = \beta_2\).}

Axiom 3 implies that for a reward sequence \(s_{0\to T-1}\), if we add a new reward \(s_T\) at the end of the sequence, then the value of the new sequence should lie between the original sequence \(s_{0\to T-1}\) and the newly added reward \(s_T\). This axiom is consistent with the experimental findings on the violation of dominance \citep{scholten2014better, jiang2017better} in intertemporal choice. To illustrate, suppose a DM is indifferent between ``receive £75 today'' (SS) and ``receive £100 in 52 weeks'' (LL). \citet{scholten2014better} find when we add a tiny reward after the payment in SS, such as changing SS to ``receive £75 today and £3 in 52 weeks'', the DM would be more likely to prefer LL over SS. \citet{jiang2017better} find the same effect can apply to LL. That is, if we add a tiny reward after the payment in LL, such as changing LL to ``receive £100 in 52 weeks and £3 in 53 weeks'', the DM would be more likely to prefer SS over LL.

Axiom 4 implies that, when the DM values a reward sequence, whether she brackets a part of its rewards into a sub-sequence should not affect how the rewards outside the sub-sequence are weighted.\footnote{If one reward is assigned a greater weight, the sequence value should be more sensitive to changes in this reward. To test this axiom, one can manipulate the framing of a sequence by bracketing some of its rewards into a sub-sequence, and then examine whether people’s valuation of the sequence becomes more (or less) sensitive to changes in a reward outside the sub-sequence. If not, this would support the axiom.} This axiom is analogous to the Independence of Irrelevant Alternatives (IIA) in discrete choice analysis. To understand its usage, we consider a three-period sequence \(s_{0\to 2} = [s_0,s_1,s_2]\), and bracket \(s_0\) and \(s_1\) into a sub-sequence \(s_{0\to 1}\). 

Suppose we can represent the value of \(s_{0\to 2}\) by \(U(s_{0\to 2}) = w_0 u(s_0) + w_1 u(s_1) + w_2 u(s_2)\), and the value of its sub-sequence by \(U(s_{0\to 1}) = \tilde{w}_0 u(s_0) + \tilde{w}_1 u(s_1)\), where for each case, the decision weights sum to one. According to Axiom 3, we can find \(\alpha \in (0,1)\) such that
\begin{equation*}
\begin{aligned}
    U(s_{0\to 2}) &= \alpha U(s_{0\to 1}) + (1-\alpha) u(s_2) \\
    & = \alpha \tilde{w}_0u(s_0) + \alpha \tilde{w}_1 u(s_1) + (1-\alpha) u(s_2)
\end{aligned}
\end{equation*}
According to Bracketing Invariance, as the bracketing operation does not involve \(s_2\), we must have \(1-\alpha = w_2\). When this condition holds, we also have \(\alpha \tilde{w}_0 = w_0\) and \(\alpha \tilde{w}_1 = w_1\). So, the decision weights for sequence \(s_{0\to 2}\) are uniquely determined. Otherwise, there may be multiple combinations of decision weights that can generate the same sequence value. 


We show in Proposition 1 that the optimal discounting model, plus Axioms 1-4, can produce the AMD model.

\noindent \textbf{Proposition 1}: \emph{Suppose \(\succsim\) admits an optimal discounting representation, then it admits an AMD representation if and only if it satisfies Axioms 1-4.}

The necessity (``only if'') is easy to see. We prove the sufficiency (``if'') in Online Appendix \ref{proof-1}. The sketch of the proof is as follows. First, consider a reward sequence \(s_{0\to T}\). Recursively applying Axioms 1 and 3 to each of its sub-sequences, we obtain decision weights \(\{w_t\}_{t=0}^T\) such that
\(s_{0\to T}\sim w_0\cdot s_0+\cdots+w_T\cdot s_T\),
with \(\sum_{t=0}^T w_t = 1\) and \(w_t>0\). Second, suppose we add a new reward \(s_{T+1}\) to the end of \(s_{0\to T}\). Adding Axioms 2 and 4, the relative weights of the first \(T+1\) periods should be invariant to the addition of \(s_{T+1}\). That is, let \(\{w'_t\}_{t=0}^{T+1}\) denote the decision weights for \(s_{0\to T+1}\), then \(\frac{w'_0}{w_0}=\frac{w'_1}{w_1}=\cdots=\frac{w'_T}{w_T}\).

Next, we look at the optimal discounting problem. The first-order condition of its solution implies that \(f'_t(w_t)=u(s_t)+\theta\), where \(\theta\) is the Lagrangian multiplier. As \(f'_t(.)\) is continuous and strictly increasing, we define a function \(\phi_t(.)\) such that
\(w_t=\phi_t(u(s_t)+\theta)\). Suppose for some real number $\Delta \theta$, \(w'_t=\phi_t(u(s_t)+\theta-\Delta \theta)\). Given that \(\frac{w'_t}{w_t}\) is invariant, we obtain that \(\frac{\phi_t(u(s_t)+\theta-\Delta \theta)}{\phi_t(u(s_t)+\theta)}\) is independent of both period \(t\) and the particular \(s_t\), and depends only on \(\Delta \theta\). This requires \(\phi_t(.)\) to take an exponential form such that \(\phi_t(u) \propto e^{u /\lambda}\), where \(\lambda > 0\). Combining this with the normalizing constraint, we can obtain the AMD representation.

\subsection{Spiking Neural Network (SNN)}

SNNs are a class of neural models that mimic key features of information processing in biological neurons \citep{maass1997networks,gerstner2002spiking,tavanaei2019deep}. In biological neural systems, neurons communicate through discrete electrical impulses, also known as spikes. A neuron is said to be active when it generates a spike. Each spike is followed by a brief refractory period. In SNNs, spike generation is typically represented through binary events: 1 for a spike, 0 for the absence of a spike. These binary events are transmitted as signals to downstream neurons, where they are integrated with signals from other neurons.

One important way to encode information in SNNs is \emph{rate coding}, whereby information is represented by each neuron's firing rate, i.e. the number of spikes generated per unit of time. Studies in neurophysiology suggest that brain encodes subjective value in a similar way \citep{sugrue2004matching,samejima2005,padoa2006neurons,padoa2013neuronal,ballesta2020values}. For example, \cite{padoa2006neurons} record neural activity in the orbitofrontal cortex of monkeys performing a juice choice task. They find that shortly after an option is presented, neurons selectively responsive to that option exhibit firing rates that increase with its value, as indicated by the amount of juice offered.

\begin{figure}[htbp]
    \centering
    \begin{subfigure}[b]{0.42\textwidth}
        \centering
        \includegraphics[width=\textwidth]{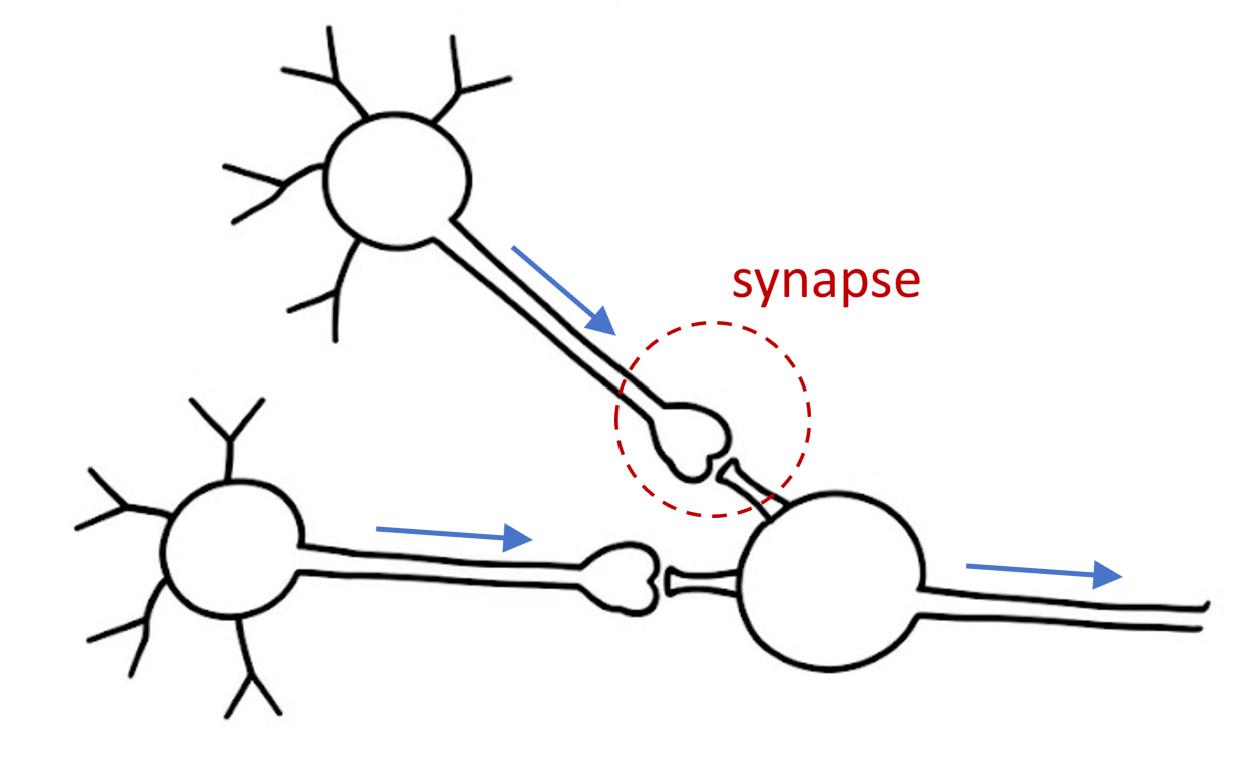}
        \caption{signal transmission across neurons}
        \label{fig:synapse}
    \end{subfigure}
    \hfill
    \begin{subfigure}[b]{0.5\textwidth}
        \centering
        \includegraphics[width=\textwidth]{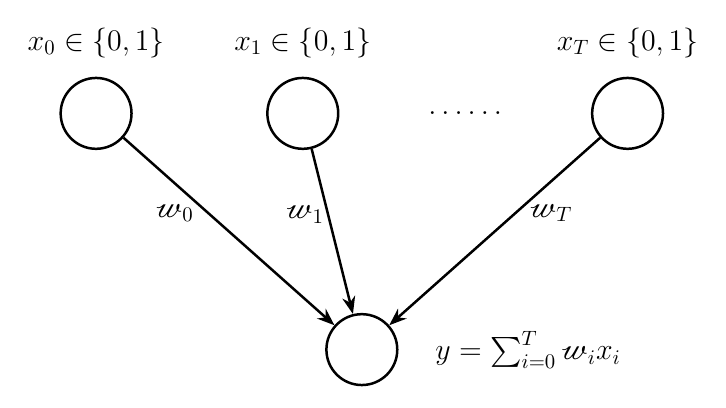}
        \caption{a two-layer SNN}
        \label{fig:simple_snn}
    \end{subfigure}
    \vspace{8pt} 

\begin{minipage}{1.0\textwidth}
{\par\footnotesize \emph{Note:} The left panel demonstrates how signals are transmitted in biological neurons. It contains two input neurons and one output neuron. Blue arrows indicate the direction of signal transmission. The connection between neurons (marked by red circles) is called a synapse, and the strength of the synapse is called the synaptic weight. 
The right panel demonstrates the SNN designed to evaluate the reward sequence \(s_{0\to T}\). Each input neuron \(i\) encodes the instantaneous utility of a reward and emits spikes at rate \(r_i\). 
At time \(\tau\), whether it fires (emits a spike) is represented by a binary event \(x_i\). The output neuron aggregates these \(x_i\) to \(y\) through a weighted sum, and the weights \(\mathscr{w_0},...,\mathscr{w_T}\) represent the synaptic weights.} 
\end{minipage}
\caption{Spiking neural network}
\end{figure}

Here, we consider a neural network of two layers: an input layer and an output layer (see Figure \ref{fig:simple_snn}). The network is designed to evaluate the reward sequence \(s_{0\to T}\). Notably, the biological brain does not necessarily process information according to this SNN. We adopt this approach just to understand the potential cognitive processes underlying the AMD model in a biologically plausible manner.

The input layer contains \(T+1\) neurons, each encoding the instantaneous utility of one reward in \(s_{0\to T}\). Let \(\tau\) denote the time elapsed as the network processes the sequence.\footnote{Note that \(\tau\) differs from \(t\). The latter indexes the reward delivery period in the sequence.} At time \(\tau\), whether input neuron \(i\) (\(i\in\{0,1,...,T\}\)) is active is indexed by \(x_i(\tau)\in\{0,1\}\). For simplicity, spike generation in the input layer is independent across neurons and over time. Input neuron \(i\) generates spikes at a fixed firing rate \(r_i\) ($0<r_i\ll 1$), such that \(E[x_i(\tau)]=r_i\). Borrowing from rate coding, we assume that \(r_i\) reflects the corresponding instantaneous utility \(u(s_i)\).

The output layer consists of a single neuron that integrates signals from the input neurons through a weighted sum.\footnote{This weighted sum can be interpreted as the membrane potential of the output neuron, or as an indicator of its instant firing probability according to some stochastic firing models \citep{gerstner2002spiking}.}  
The weight associated with each signal reflects the strength of the connection between the output neuron and the neuron generating that signal. In neuroscience, such weights are called the synaptic weights (see Figure \ref{fig:synapse}).

The synaptic weights are dynamic. According to \citet{turrigiano2000hebb}, their dynamics are driven by two fundamental principles: Hebbian learning and homeostatic scaling. Hebbian learning is a positive feedback mechanism---when two connected neurons are active together, their synaptic weight tends to increase, thereby enhancing communication between them. \cite{hebb1949} summarizes this principle as ``neurons that fire together, wire together''.\footnote{
 Some variants of this principle, such as spike-timing-dependent plasticity \citep{song2000competitive}, emphasize that the temporal order of spikes between connected neurons also influences the dynamics of synaptic weights. For simplicity, we abstract away from this temporal dependence.} Homeostatic scaling is a negative feedback mechanism---when the integrated signal becomes too strong, the output neuron downscales all its synaptic weights by a common factor to weaken communication with input neurons, and vise versa. This maintains the activity of the output neuron within a stable and functional range \citep{royer2003conservation,turrigiano2008self,ibata2008rapid}.

We formalize these two principles as follows. Let  \(\mathscr{w_i}(\tau)\) denote the synaptic weight between the output neuron and neuron \(i\) at time \(\tau\).
We assume that, at each time \(\tau\), the output neuron integrates all inputs to an output \(y(\tau) = \sum_{i=0}^T \mathscr{w_i}(\tau) x_i(\tau)\) and updates \(\mathscr{w_i}(\tau)\) in two steps. First, the synaptic weight is updated to \(\tilde{\mathscr{w}}_i(\tau)\) according to a variant of Hebbian learning known as the covariance rule \citep{gerstner2002spiking}. That is, \(\tilde{\mathscr{w}}_i(\tau)=\mathscr{w_i}(\tau) + \Delta \tilde{\mathscr{w}}_i(\tau)\), and the change \(\Delta \tilde{\mathscr{w}}_i\) is given by
\[\label{eq:hebbian_rule}
\Delta \tilde{\mathscr{w}}_i(\tau) = \eta (x_i(\tau) - r_i)(y(\tau)-\bar{y}(\tau))
\]
where $\eta$ denotes the learning rate ($0< \eta \ll1$), and $\bar{y}(\tau) = \sum_{i=0}^T \mathscr{w_i}(\tau)r_i$. Equation (\ref{eq:hebbian_rule}) implies that the expected change in \(\mathscr{w_i}\) is proportional to the covariance between \(x_i(\tau)\) and \(y(\tau)\). Intuitively, the synaptic weight tends to increase when the activities of the two neurons are more positively correlated. Second, each synaptic weight is multiplied by a common factor equal to the reciprocal of their sum, yielding \(\mathscr{w_i}(\tau+1) = \frac{\tilde{\mathscr{w}}_i(\tau)}{\sum_{j=0}^T \tilde{\mathscr{w}}_j(\tau)}\). This provides a simple form of homeostatic scaling that maintains the output \(y(\tau)\) between 0 and 1. The use of this normalization form can be traced back to \cite{rochester1956tests}. Also, it keeps the total synaptic weight constant, which is consistent with the finding of \cite{royer2003conservation}. 

Overall, Hebbian learning provides a rationale for assigning greater decision weights to larger rewards, and homeostatic scaling provides a rationale for the normalizing constraint. In Proposition 2, we show that the AMD factors can be approximated by the expected dynamics of synaptic weights in this SNN.

\noindent \textbf{Proposition 2:} \emph{If $\eta r_{\max}^2 + \eta^2 r_{\max}(1 + \sum_{j=0}^Tr_j)^2 \ll \frac{1}{\tau}$, the expected synaptic weight between the output neuron and input neuron $i$ at time $\tau$ can be written as
\[\label{eq:snn_weight_approx}
E[\mathscr{w_i}(\tau)] = \frac{\mathscr{w_i}(0) e^{\eta\tau r_i}}{\sum_{j=0}^T \mathscr{w}_j(0) e^{\eta\tau r_i}} + R_i(\tau)
\]
where $R_i(\tau)$ is an error of order $O(\eta \tau r_{\max}^2 + \eta^2\tau r_{\max}(1 + \sum_{j=0}^Tr_j)^2)$, $\mathscr{w_i}(0)$ is the initial weight at $\tau=0$, and $r_{\max}$ is the highest firing rate among all input neurons.}

Every element of the AMD factors has a counterpart in Equation (\ref{eq:snn_weight_approx}): $\frac{1}{\lambda}=\eta \tau$, $d_t = \mathscr{w}_t(0)$, and $u(s_t)$ is reflected by $r_t$. This correspondence suggests that our proposed SNN provides a neuro-computational mechanism that can approximately characterize the AMD model. Using Proposition 2 and the mean-field approximation, it follows readily that the expected output $E[y(\tau)]$ in this SNN is approximately equal to the sequence value $U(s_{0\to T})$ in the AMD model. Therefore, the dynamics of the output \(y(\tau)\) could represent the process by which the SNN learns the sequence value. Suppose we determine a time at which this SNN stops updating its weights, and from then on we keep recording its outputs. Then the average recorded output should approximate the sequence value obtained from the AMD model under these corresponding parameters.

Importantly, the approximation is valid only under the inequality condition at the beginning of Proposition 2. That condition implies that both the cumulative firing rates $\sum_{j=0}^T r_j$ and the processing time $\tau$ cannot be too large; otherwise, the error term $R_i(\tau)$ would no longer be negligible relative to the leading term.\footnote{To achieve that condition, one can set the number of rewards with non-zero utilities to at most $1/r_{\max}$ and set the stopping time of weight updating to be on the order of $1/(\eta r_{\max})$. Alternatively, the stopping time can be determined endogenously. For instance, note that the synaptic weights tend to concentrate over time on the input neuron with the highest firing rate, which increases $E[y(\tau)]$. We can therefore tradeoff the increase of $E[y(\tau)]$ (as the approximate sequence value) against the time cost of updating.}

We prove Proposition 2 in Online Appendix \ref{proof-2}. The sketch of the proof is as follows. Let $\mathscr{W}(\tau)$ denote the synaptic weights achieved at time $\tau$. Since spike generation is independent, by taking the expectation of Equation \eqref{eq:hebbian_rule}, we get \(E[\Delta \tilde{\mathscr{w}}_i(\tau) | \mathscr{W}(\tau)] = \eta \mathscr{w_i}(\tau) \text{Var}(x_i)\), where \(\text{Var}(x_i) = r_i(1-r_i)\). This implies that \(E[ \tilde{\mathscr{w}}_i(\tau) | \mathscr{W}(\tau)] = \mathscr{w_i}(\tau)\cdot [1+\eta \text{Var}(x_i)]\).
There are three key approximations in the proof. First, when \(r_i\) is small, we can use \(r_i\) to approximate \(\text{Var}(x_i)\). Second, when \(\eta\) is small, we can use \(e^{\eta r_i}\) to approximate \(1+ \eta r_i\). Third, we can use \(\frac{E[\tilde{\mathscr{w}}_i]}{\sum_{j=0}^T E[\tilde{\mathscr{w}}_j]}\) to approximate \(E\left[\frac{\tilde{\mathscr{w}}_i}{\sum_{j=0}^T \tilde{\mathscr{w}}_j}\right]\). Recursively applying these approximations to each time step of weight updating, we can obtain Equation \eqref{eq:snn_weight_approx}.

\section{Behavioral Implications}\label{sec:implications}

\subsection{Hidden Zero Effect}\label{hidden-zero-effect}

Empirical evidence suggests that time discounting depends on how sequences are framed. A well-known example is the hidden zero effect \citep{magen2008hidden}. Recent studies typically relate this effect to attention mechanisms \citep{radu2011mechanism,read2017value,chen2026attentional,kang2026relationship}.

To illustrate the hidden zero effect, suppose the DM is indifferent between ``receive £100 today'' (SS) and ``receive £120 in 25 weeks'' (LL). The effect implies that people exhibit more patience when SS is framed as a sequence rather than as a single-period reward. In other words, if we frame SS as ``receive £100 today and £0 in 25 weeks'' (SS1), the DM would prefer LL to SS1. Similar to the violation of dominance, this phenomenon can be viewed as a justification for the Sequential Outcome-Betweenness axiom in Section \ref{optimal-discounting}. Moreover, this effect is asymmetric \citep{read2017value}. That is, framing LL as ``receive £0 today and £120 in 25 weeks'' (LL1) has no effect on preference.

The AMD model explains the hidden zero effect as follows. When the DM evaluates a reward sequence \(s_{0\rightarrow T}\), she divides a fixed amount of attention over \(T\) periods. In the given example, the DM may perceive the sequence length of SS as ``today'' and the length of SS1 as ``25 weeks''. For option SS, she focuses on the current period in which she can get £100. For option SS1, she pays some of her attention to the future periods in which no reward is delivered, and this reduces the decision weight assigned to the current period. As a result, she values SS1 lower than SS. By contrast, the DM may perceive both LL and LL1 as having a sequence length of “25 weeks,” since both deliver zero rewards from the current period until just before “25 weeks”. According to our model, when she evaluates LL, she has already paid some attention to all periods earlier than ``25 weeks''. So, changing LL to LL1 does not affect her choice.

\subsection{Relation to Hyperbolic Discounting}\label{relation-to-hyperbolic-discounting}

Most studies on intertemporal choice involve only comparisons between single-period rewards (SS and LL). Here, we derive the AMD factor for SS/LL and use it to illustrate how our model can account for anomalies in this setting. For simplicity, we assume the DM's default discount factor is exponential, so that she is time-stationary under the initial status; that is, \(d_t = \delta^t\), where \(\delta\in(0,1]\).


Consider a reward sequence \(s_{0\rightarrow T}\) in which \(u(s_t)=0\) for all \(t<T\), and \(u(s_T)>0\). This implies the DM receives no reward until period \(T\). In this case, the DM's valuation of \(s_{0\rightarrow T}\) is given by \(U(s_{0\rightarrow T})=w_Tu(s_T)\). Let \(v(x)=u(x)/\lambda\). By Definition 1, we can derive \(w_T\) as a function of \(s_T\):
\[\label{eq:w_T} 
w_T = \frac{1}{1+G(T)e^{-v(s_T)}} 
\]
where
\begin{equation*}
    G(T) = \left\{ 
    \begin{aligned} 
    & \frac{1}{1-\delta}(\delta^{-T}-1) \; , & 0<\delta<1\\ 
    & T\; ,    & \delta=1\ 
    \end{aligned} \right. 
\end{equation*} 
In Equation \eqref{eq:w_T}, \(w_T\) can represent the discount function for single-period reward \(s_T\). Interestingly, when \(\delta=1\), \(w_T(s_T)\) takes a form similar to hyperbolic discounting. 

In recent years, several studies have attempted to provide a rational account of hyperbolic discounting, including by aggregating heterogeneous discount rates \citep{gollier2005aggregation,ebert2020weighted} and by Bayesian learning \citep{gershman2020rationally,gabaix2022myopia}. Our model differs from them as it achieves this by normalizing discount factors across periods. In terms of functional form, the closest work to ours is  \cite{gershman2020rationally}, which can be viewed as a variant of the AMD model under certain specifications.\footnote{\citet{gershman2020rationally} propose that the discount function for a single-period reward \(s_T\) is \(\frac{1}{1+(\gamma s_T)^{-1}T}\), where \(\gamma>0\). In Equation (\ref{eq:w_T}), if we set \(\delta=1\) and \(v(s_T)=\ln(\gamma s_T+1)\), we will obtain \(w_T = \frac{1}{1+(\gamma s_T+1)^{-1}T}\).}.

In the following three subsections, we use Equation (\ref{eq:w_T}) to explain three decision anomalies: the common difference effect (and its reverse), risk aversion over time lotteries, and S-shaped value function.

\subsection{Common Difference Effect}\label{common-difference-effect}

The common difference effect \citep{loewenstein1992anomalies} implies that, when the DM faces a choice between LL and SS, adding a common delay to both options can increase her preference for LL. \citet{sayman2009investigation} document that this effect can reverse: the preference for LL sometimes decreases after a common delay is added.

Let \((v_l,t_l)\) denote a reward of utility \(v_l\) delivered at period \(t_l\), and \((v_s,t_s)\) denote a reward of utility \(v_s\) delivered at period \(t_s\). We set \(v_l>v_s>0\), \(t_l>t_s>0\). So, \((v_l,t_l)\) can represent a LL and \((v_s,t_s)\) can represent a SS. We denote the discount factors for LL and SS by \(w_{t_l}(v_l)\) and \(w_{t_s}(v_s)\). Suppose \(w_{t_l}(v_l)\cdot v_l = w_{t_s}(v_s)\cdot v_s\). The common difference effect implies that \(w_{t_l+\Delta t}(v_l)\cdot v_l > w_{t_s+\Delta t}(v_s)\cdot v_s\), where \(\Delta t >0\). For the discount function specified in Equation (\ref{eq:w_T}), we can obtain Proposition 3.

\noindent \textbf{Proposition 3:} \emph{The following statements are true for AMD:
\begin{enumerate} \def\labelenumi{(\alph{enumi})}
\item If \(\delta=1\), the common difference effect always holds.
\item If \(0<\delta<1\) (the DM is initially impatient), the common difference effect holds when and only when \(v_l-v_s+\ln(v_l/v_s)>(t_l-t_s)\ln(1/\delta)\).
\end{enumerate}
}

\cite{sayman2009investigation} find that the common difference effect reverses more often when \(t_s\) and \(t_l\) are shorter.\footnote{Based on this observation, \cite{sayman2009investigation} propose that the discount function may be concave for short delays and convex for long delays. We discuss this in the next subsection.} In contrast, statement (b) in Proposition 3 yields a novel prediction: for an impatient DM, the common difference effect holds only if the relative and absolute differences in reward utility between LL and SS are significantly larger than the absolute difference in time delay; otherwise, a reverse common difference effect may be observed.


The proof of Proposition 3 is in Online Appendix \ref{proof-3}. We outline the underlying intuition as follows. When the DM is impatient, adding a common delay would naturally make \(v_l\) and \(v_s\) more discounted; so, less attention is paid to them. Suppose the sum of decision weights is not changed. Then DM can ``free up'' some attention that is originally captured by these two rewards and reallocate it to other periods in each option. There are three mechanisms jointly determining whether we could observe the common difference effect under this circumstance.

First, in each option, the existing periods which have no reward delivered grab some attention. So, the DM would assign more weights to some zero-utility rewards, delivered in duration \([0,t_l)\) for LL and in duration \([0,t_s]\) for SS. Given \(t_l >t_s\), the duration in LL would naturally capture more attention than the duration in SS. In other words, adding the common delay makes the DM focus more on the original waiting time in LL than in SS, which decreases her preference for LL.

Second, the newly added delay grabs some attention. The DM would assign some weights to rewards (of zero utility as well) delivered in duration \((t_l,t_l+\Delta t)\) for LL and in duration \((t_s, t_s+\Delta t)\) for SS. For LL, there have already been a lot of periods that capture the DM's attention, before adding the delay \(\Delta t\). Thus, the duration \((t_l,t_l+\Delta t)\) in LL captures less attention than its counterpart in SS. This increases the DM's preference for LL.

Third, the only positive reward, delivered in \(t_l\) for LL and in \(t_s\) for SS, could draw some attention back. Given that the DM attends more to larger rewards, the positive reward in LL can capture more of the ``freed-up'' attention than that in SS. This also increases the preference for LL. If the latter two mechanisms override the first mechanism, we would observe the common difference effect.

\begin{figure}[h]
\vspace{12pt}
\centering 
\includegraphics[width=0.64\textwidth]{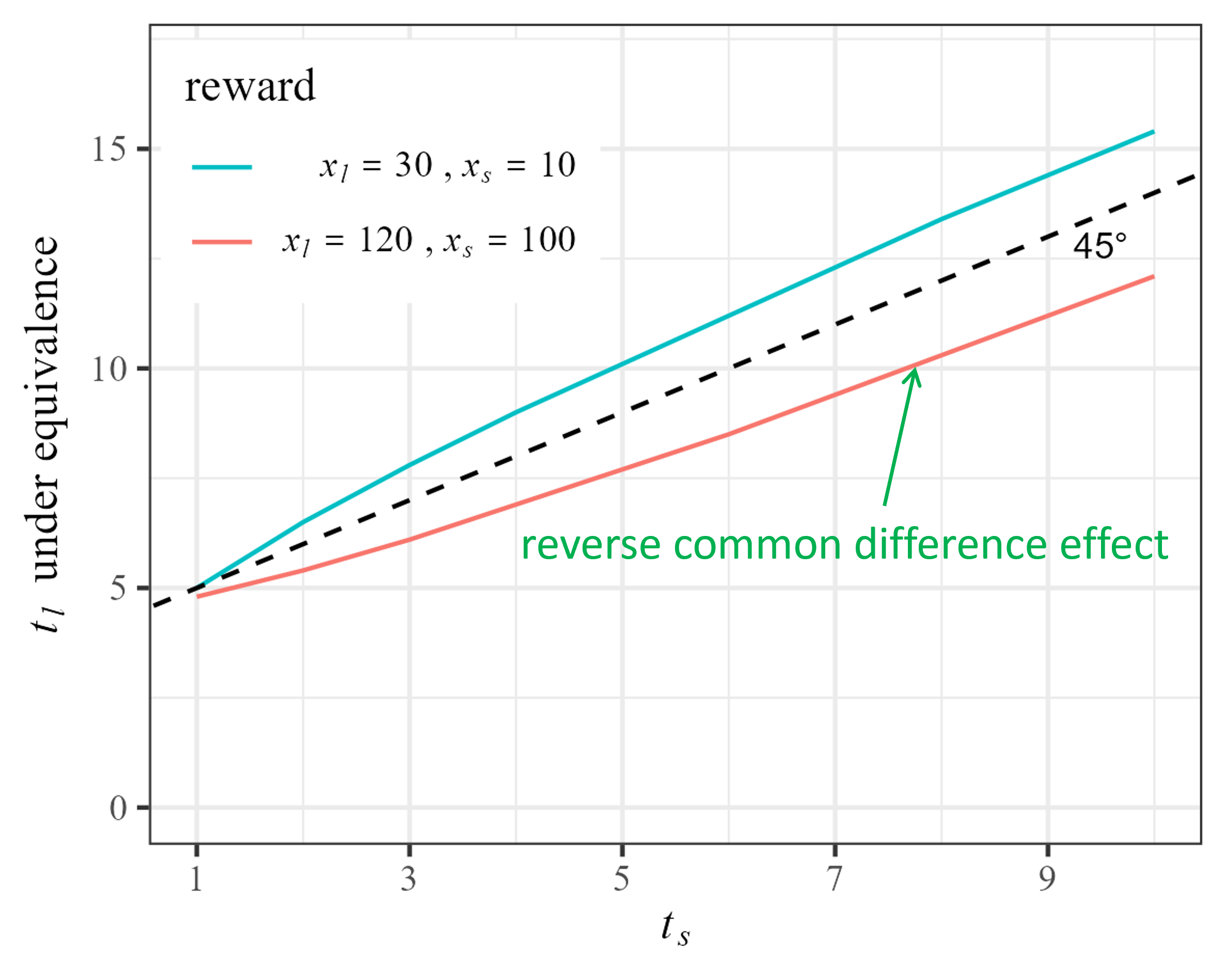}
\vspace{8pt} 

\begin{minipage}{1.0\textwidth} {\par\footnotesize Note: $x_l$ and $x_s$ are the positive reward levels for LL and SS. The values of LL and SS are calculated based on Equation (4). $d_t=0.75^t$, $u(x)=x^{0.6}$, $\lambda=2$. For each level of $t_s$, we identify the delay $t_l$ that makes the value of LL equivalent to SS. The blue line (above) demonstrates the common difference effect, and the red line (below) demonstrates the reverse common difference effect.} 
\end{minipage} 
\caption{The common difference effect and its reverse} 
\label{fig:common_diff} 
\end{figure}

Figure \ref{fig:common_diff} demonstrates an example for the reverse common difference effect. For each level of \(t_s\), we identify the \(t_l\) that makes the value of LL equivalent to SS. If the common different effect is valid, increasing \(t_s\) and \(t_l\) by the same level would make the DM prefer LL. Under this condition, for one unit increase in \(t_s\), to make LL and SS valued equally, the identified \(t_l\) should be increased by a time greater than one unit. On the contrary, if the reverse common difference effect is valid, for one unit increase in \(t_s\), the identified \(t_l\) should be increased by a time smaller. In Figure \ref{fig:common_diff}, the blue line (above) reflects the common different effect, while the red line (below), which has a lower \(v_l-v_s\) and \(v_l/v_s\), reflects the reverse of it.

\subsection{Concavity of Discount Function}\label{concavity-of-discount-function}

Many time discounting models, such as exponential and hyperbolic discounting, assume the discount function is convex in time delay. However, experimental evidence suggests that the discount function actually used in people's decision-making may be inverse-S-shaped. That is, it may be concave for short delays and convex for long delays \citep{onay2007intertemporal,sayman2009investigation,takeuchi2011non}. In Proposition 4, we show that the AMD model produces this shape of the discount function when the DM is impatient and the reward amount is large enough.

\begin{figure}[h] 
\vspace{12pt} 
\centering 
\begin{subfigure}{0.49\textwidth}
\centering 
\includegraphics[width=\linewidth]{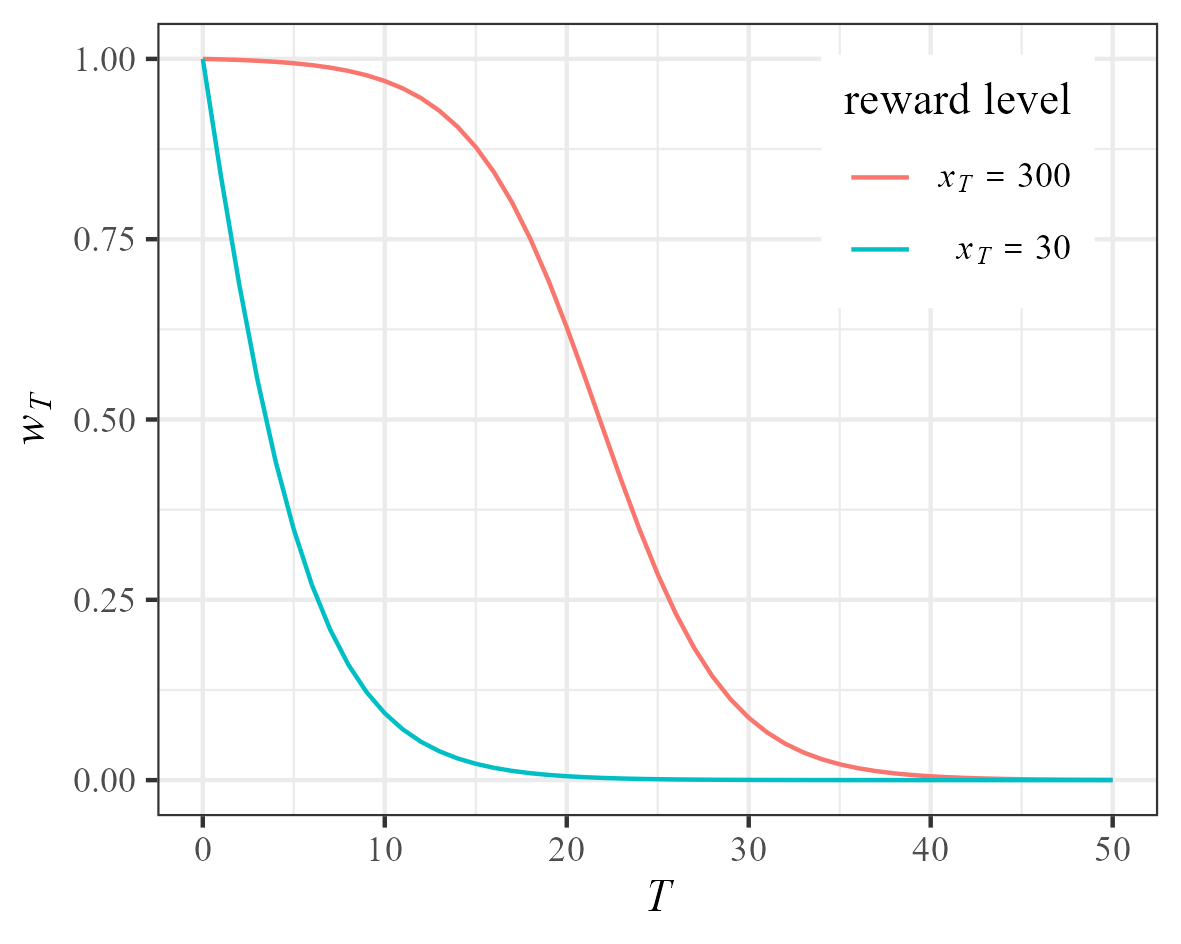} 
\caption{discount function} 
\label{fig:concave_discount}
\end{subfigure} 
\hfill 
\begin{subfigure}{0.49\textwidth} 
\centering 
\includegraphics[width=\linewidth]{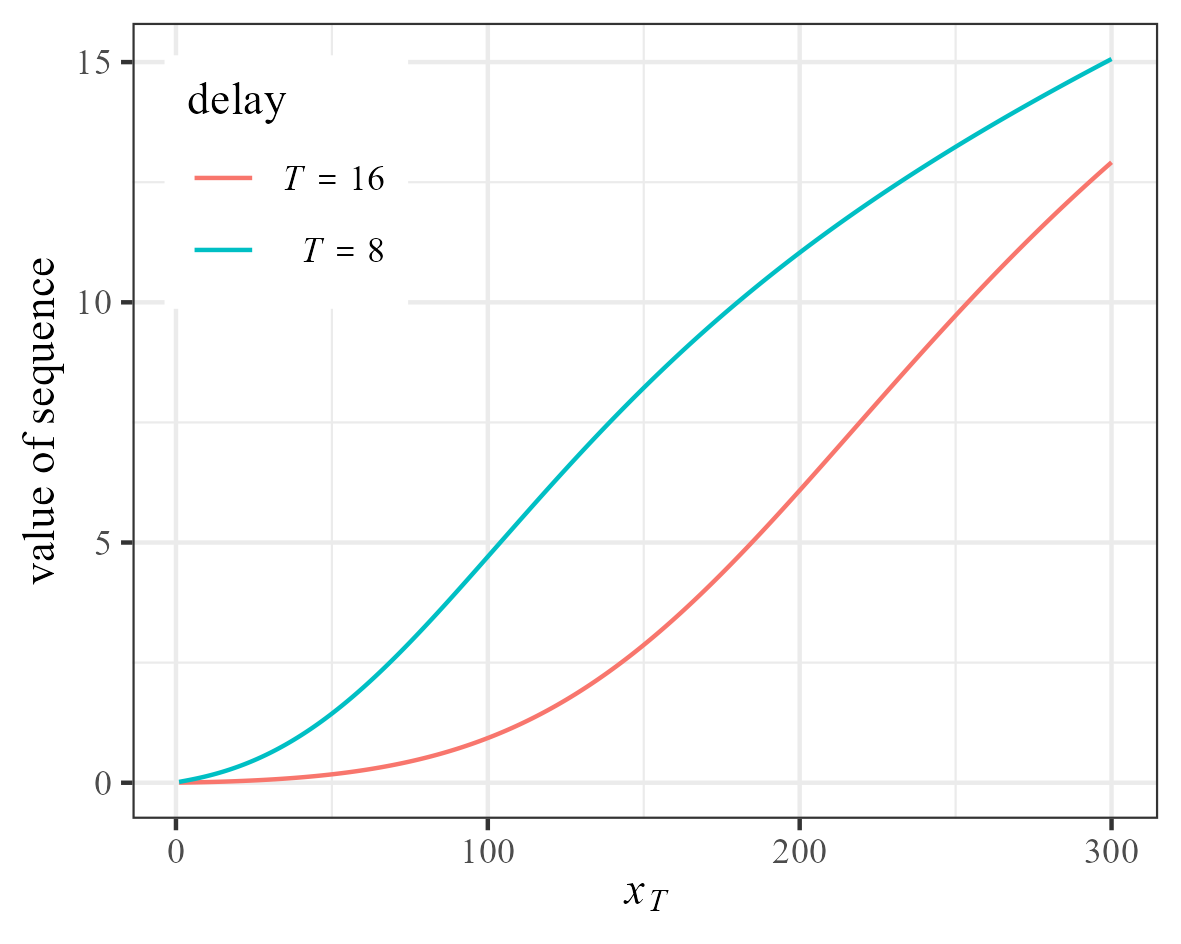}
\caption{value function}
\label{fig:s_shape_value}
\end{subfigure} 
\vspace{8pt} 

\begin{minipage}{1.0\textwidth} 
{\par\footnotesize Note: A reward of amount $x_T$ is delivered at period $T$. The discount function and value function are calculated based on Equation (4). $d_t=0.75^t$, $u(x)=x^{0.6}$, $\lambda=2$.} 
\end{minipage}
\caption{Discount function and value function for a delayed reward}
\label{fig:discount_value_function} 
\end{figure}

\noindent \textbf{Proposition 4:} \emph{Suppose a single-period reward of amount \(X\) is delivered at period \(T\). Let \(w_T\) denote the AMD factor for this reward. If \(\delta =1\), then \(w_T\) is convex in \(T\). If \(0<\delta<1\), there exists a reward threshold \(\underline{X}>0\) and a time threshold \(\underline{T}>0\) such that:
\begin{enumerate} \def\labelenumi{(\alph{enumi})}
\item when \(X\leq \underline{X}\), \(w_T\) is convex in \(T\);
\item when \(X > \underline{X}\), \(w_T\) is convex in \(T\) given \(T\geq \underline{T}\), and it is concave in \(T\) given \(0<T<\underline{T}\).
\end{enumerate}
}
The proof of Proposition 3 is in Online Appendix \ref{proof-4}. Figure \ref{fig:concave_discount} demonstrates the convex discount function (blue line, below) and the inverse-S shaped discount function (red line, above) obtained from Equation \eqref{eq:w_T}.

The inverse-S-shaped discount function helps explain people's risk attitudes toward time lotteries. To illustrate, consider a choice between a risky option and a sure option. In the risky option, a reward is delivered at period \(t_l\) with probability \(\pi\) and at period \(t_s\) with probability \(1-\pi\), where \(0<\pi<1\). In the sure option, the same reward is delivered at period \(t_m\), where \(t_m=\pi t_l +(1-\pi) t_s\). The convex discount function predicts that the DM is always risk seeking over time lotteries \citep{dejarnette2020time}. That is, the DM would prefer the risky option to the sure option. For example, she would prefer receiving an amount of money today or in 20 weeks with equal probability, rather than receiving it in 10 weeks with certainty. However, \cite{onay2007intertemporal} find that, when \(\pi\) is small, people are risk averse over time lotteries. Since \(t_m\) is increasing in \(\pi\), the observed risk aversion can be explained by the discount function being concave for short delays.

\subsection{S-Shaped Value Function}\label{s-shaped-value-function}

In decision theories, a common assumption about the utility function \(u(.)\) is \(u''<0\). Usually, this yields that the value function of a reward is concave. However, empirical evidence suggests that the value functions are often S-shaped. Such S-shaped value functions can be generated by various approaches, such as reference dependence \citep{kahneman1979prospect} and efficient coding of numbers \citep{louie2012efficient}. Through the AMD model, we provide a novel account for S-shaped value function based on the insight that larger rewards capture more attention.

Consider a single-period reward of amount \(X\) delivered at period \(T\). Its value function can be represented by \(U(X,T)=w_T(X)u(X)\). We assume \(u'>0\), \(u''<0\), and \(w_T\) is determined by Equation (\ref{eq:w_T}). \(w_T\) is increasing in \(X\) as the DM tends to pay more attention to larger rewards. Both functions \(u(X)\) and \(w_T(X)\) are concave in \(X\); so when \(X\) is small, they both grow fast. At some conditions, it is possible that the product of the two functions is convex in \(x\) when \(x\) is small enough. We derive the conditions for the S-shaped value function in Proposition 5.

\noindent \textbf{Proposition 5:} \emph{Suppose \(T\geq1\), and \(\frac{d}{dX}\left(\frac{1}{v'(X)}\right)\) is continuous in \((0,+\infty)\), then:
\begin{enumerate} \def\labelenumi{(\alph{enumi})}
\item There exists a threshold \(\bar{X} \in\mathbb{R}_{\geq0}\) such that \(U(X,T)\) is strictly concave in \(X\) when \(X\in [\bar{X},+\infty)\).
\item If \(\lim_{X\to 0^+}\frac{d}{dX}\left(\frac{1}{v'(X)}\right)<1\), there exists \(X^*\in(0, \bar{X})\) such that, for any \(X\in (0,X^*)\), \(U(X,T)\) is strictly convex in \(X\).
\item There exists a threshold \(\lambda^*\) and an interval \((X_1,X_2)\) such that, if \(\lambda<\lambda^*\), for any \(X\in(X_1,X_2)\), \(U(X,T)\) is strictly convex in \(X\), where \(\lambda^*>0\) and \((X_1,X_2)\subset(0,\bar{X})\).
\end{enumerate}
}

The proof of Proposition 5 is in Online Appendix \ref{proof-5}. Proposition 5 implies, if the derivative of \(\frac{1}{v'(X)}\) converges to a small number when \(X\rightarrow 0^+\), or the cost of attention reallocation \(\lambda\) is small enough, the value function \(U(X,T)\) will be S-shaped in some interval of \(X\).\footnote{For the CRRA utility function \(u(X) = X^{1-\gamma}/(1-\gamma)\), the former condition requires that \(\gamma>1\).} Figure \ref{fig:s_shape_value} demonstrates two examples of this S-shaped value function.

\subsection{Intertemporal Correlation Aversion}\label{intertemporal-correlation-aversion}

Consider a DM facing two lotteries, \(L1\) and \(L2\). For lottery \(L1\), she can receive £100 today and £100 in 30 weeks with probability 1/2, and receive £3 today and £3 in 20 weeks with probability 1/2. For lottery \(L2\), she can receive £3 today and £100 in 30 weeks with probability 1/2, and receive £100 today and £3 in 20 weeks with probability 1/2. In lottery \(L1\), rewards delivered at the two different periods are positively correlated; in lottery \(L2\), those rewards are negatively correlated. The expected discounted utility theory predicts the DM is indifferent between the two lotteries. However, recent studies find the evidence of \emph{intertemporal correlation aversion} \citep{andersen2018multiattribute, rohde2023intertemporal}. That is, people usually prefer lottery \(L2\) to \(L1\).\footnote{For theoretical analysis about intertemporal correlation aversion, please see \citet{epstein1983stationary}, \citet{epstein1989substitution}, \citet{weil1990nonexpected}, \citet{bommier2007}, and \citet{bommier2017monotone}. The AMD model takes a similar form to the class of models defined by \citet{epstein1983stationary}. A key feature of such models is that the discount factor for future utilities is dependent on the utility achieved in the current period.}

The above example can be explained by the AMD model as follows. When evaluating a lottery, the DM first values the sequence received under each state and then aggregates the values to calculate the certainty equivalent. For simplicity, suppose there are only two periods. Note when all rewards in a sequence are equal, the AMD factor for every period \(t\) in the sequence remains the same as its default discount factor. So, for lottery \(L1\), the decision weight assigned to each period is constant across states. Suppose she allocates decision weight \(w\) to the first period and \(1-w\) to the second period. 

For lottery \(L2\), in the state that the DM can receive £100 in the first period and £3 in the second period, the reward £100 can capture more attention so that its decision weight, say \(w'\), is greater than \(w\). Similarly, in the other state, the decision weight for the later reward £100, say \(1-w''\), is greater than \(1-w\). Therefore, the value of \(L1\) can be represented by \(0.5\cdot u(3)+0.5\cdot u(100)\), and the value of \(L2\) can be represented by \(0.5(1-w'+w'')\cdot u(3)+0.5(1-w''+w')\cdot u(100)\). Given \((1-w'')+w'>1-w+w=1\), the value of \(L2\) is greater than that of \(L1\).

In a more general setting, whether the AMD model can robustly produce intertemporal correlation aversion is influenced by \(\lambda\). To see this, we adopt the definition of intertemporal correlation aversion in \citet{bommier2007}. Let \((s_1,s_2)\) denote a state in which the DM can receive reward \(s_1\) in period \(t_1\) and then reward \(s_2\) in period \(t_2\), where \(t_2>t_1\geq 0\). Lottery \(L1\) generates \((X_s,Y_s)\) and \((X_l,Y_l)\) with equal probability, while lottery \(L2\) generates \((X_s,Y_l)\) and \((X_l,Y_s)\) with equal probability, with \(X_l>X_s>0\), \(Y_l>Y_s>0\). Both lotteries have the same \(t_1\) and \(t_2\), and the same sequence length. By Proposition 6, we show that in this setting, we can always find a \(\lambda\) that makes the DM intertemporal correlation averse.

\noindent \textbf{Proposition 6:} \emph{Suppose \(U(L1), U(L2)\) are the values of lotteries \(L1\) and \(L2\) calculated based on the AMD model. For any \(X_l>X_s>0\), \(Y_l>Y_s>0\), any sequence length of lotteries, and any default discount factors, there exists a threshold \(\lambda^{**}\) such that for all cost of attention reallocation \(\lambda\in(\lambda^{**},+\infty)\), we have \(U(L1)<U(L2)\), i.e.~the DM exhibits intertemporal correlation aversion.}

The proof of Proposition 6 is in Online Appendix \ref{proof-6}. The threshold \(\lambda^{**}\) is jointly determined by \(X_l\), \(Y_l\), \(Y_s\), as well as the default discount factors for periods \(t_1\) and \(t_2\). Notably, when \(\lambda \leq \lambda^{**}\), the DM may be intertemporal correlation seeking under some conditions.\footnote{To validate, one can set \(u(X_s)=5\), \(u(X_l)=10\), \(u(Y_s)=1\), \(u(Y_l)=3\). Suppose the sequence offered by each lottery contains only two periods, \(t_1\) and \(t_2\), and the default discount factors are uniformly distributed, i.e.~\(d_{t_1}=d_{t_2}\). In this case, setting \(\lambda=1\) would generate intertemporal correlation seeking, while setting \(\lambda=100\) would generate intertemporal correlation aversion.} This suggests a potentially new mechanism for intertemporal correlation aversion; that is, the DM exhibits intertemporal correlation aversion because the process by which she shifts attention to larger rewards is very costly.

\subsection{Concentration Bias}\label{concentration-bias}

\cite{dertwinkel2022concentration} document a phenomenon called concentration bias. Consider a DM with a consumption budget £100 to spend over four days (from period 0 to period 3). Suppose the DM has two options: concentrating all consumption into period 0, or splitting the consumption evenly over four periods. The concentration bias would make her prefer the first option to the second option. According to \cite{kHoszegi2013model}, this phenomenon can arise when the decision weight assigned to a reward increases with its amount (or with the deviation of that amount from a reference point).  

Following the example, we denote the first option as sequence {[}100,0,0,0{]} and the second option as sequence {[}25,25,25,25{]}. We show that the AMD model can produce concentration bias in the same way as predicted by \cite{kHoszegi2013model}. For convenience, the default discount factor for any period \(t\) is \(d_t = \delta^t\) and \(0<\delta<1\). According to the AMD model, the DM prefers the first option if and only if
\begin{equation*}
    \frac{e^{u(100)/\lambda}}{e^{u(100)/\lambda}+\delta+\delta^2+\delta^3}\cdot u(100)>u(25) 
\end{equation*}
Obviously, this inequality holds only when \(\delta\) or \(\lambda\) is small enough, which implies that the DM should be very impatient or can reallocate attention at a very low cost. Notably, it is also possible that the DM prefers concentrating all consumption into the final period rather than evenly splitting the consumption. In this case, the decision weight multiplied by \(u(100)\) in the inequality would be \(\frac{\delta^3 \exp\{u(100)/\lambda\}}{1+\delta+\delta^2+\delta^3 \exp\{u(100)/\lambda\}}\). Then, the inequality holds only when both \(\delta\) is large enough and \(\lambda\) is small enough. Both cases are in line with the claim in \citet{kHoszegi2013model} and \citet{dertwinkel2022concentration} that concentration bias can make people behave too impatiently or too patiently.

The above example provides the intuition that, for concentration bias to arise, the cost of attention reallocation \(\lambda\) must be sufficiently small. We formalize this intuition as follows. Suppose the DM has a consumption budget \(m\) (\(m>0\)) to spend over \(T\) periods. Let reward sequence \(s_{0\rightarrow T}\) represent her consumption plan at period 0, and let \(A\subset \mathbb{R}_{\geq 0}^{T+1}\) denote her option space. In period 0, DM wants to find a \(s_{0\rightarrow T}\) that solves the optimization problem:
\[\label{eq:max_consumption}
\max_{s_{0\rightarrow T}\in A}\;\sum_{t=0}^T w_t u(s_t) 
\]
where
\begin{equation*}
    A=\left\{s_{0\rightarrow T} \bigg|\;\sum_{t=0}^T s_t = m,\; \forall t:s_t \geq 0\right\}
\end{equation*}
and \(w_t\) is the AMD factor for consumption in period \(t\). For \(s\in[0,m]\), we have \(u'(s)>0\), \(u''(s)<0\). 

We denote the optimization problem in Equation (\ref{eq:max_consumption}) by \(\mathcal{O}(m,A,\{d_t\}_{t=0}^T)\). By Proposition 7, when the DM is impatient and the cost of attention reallocation \(\lambda\) is small enough, her optimal consumption plan over the entire option space is to consume all of \(m\) immediately, provided that her utility function satisfies a certain condition.\footnote{This condition is satisfied, for example, by the logarithmic utility function $u(s)=\ln(1+\gamma s)$ and the CARA utility function $u(x)=1-e^{-\gamma x}$, where $\gamma>0$.} The proof of Proposition 7 is in Online Appendix \ref{proof-7}. In Proposition 7, the conditions specified for the emergence of concentration bias are not explicitly identified in \citet{kHoszegi2013model}. 

\noindent \textbf{Proposition 7}: \emph{Suppose the DM faces the planning problem \(\mathcal{O}(m,A,\{d_t\}_{t=0}^T)\), and for any period \(t<T\), we have \(d_t >d_{t+1}>0\). If \(\frac{d}{ds}\left(\frac{1}{u'(s)}\right)\) has an upper limit for \(s \in[0,m]\), then there exists a threshold \(\underline{\lambda}>0\) such that for any \(\lambda\leq\underline{\lambda}\), her optimal consumption plan is to concentrate all consumption into period 0.}



\subsection{Inconsistent Planning and Learning}\label{inconsistent-planning-and-learning}

Numerous studies suggest that people perform time-inconsistent behaviors in their daily lives \citep{ericson2019intertemporal}. For example, they often consume more than they originally planned and procrastinate on effortful tasks. In a general sense, such behaviors can be termed present-biased behaviors. Several theories have been proposed to explain such behaviors, such as dual-system preferences \citep{laibson1997golden}, naivete \citep{o1999doing}, and reference dependence \citep{kHoszegi2009reference}. Based on the AMD model, we provide an alternative explanation for these behaviors. That is, in dynamic decision-making, people update their default discount factors over time. 

In intuition, during each decision period, people may draw on their past experiences when they allocate attention. If, in the last step, they allocate much attention to a particular period, they may then take this as a given or default status in the following step and continue to add attention to it. This yields present-biased behavior. Compared with existing theories, our explanation is built on learning and memory. To perform present-biased behaviors, the DM should recall her mental status from the last step and allocate attention accordingly. A memoryless DM may perform the reverse behavior. We analyze this using the consumption planning problem defined in the last subsection.

Again, we suppose the DM has a budget \(m\) for consumption. She needs to allocate it over \(T\) periods (\(T\geq 2\)) and the end of sequence is a fixed date. In period 0, her default discount factors \(\{d_t\}_{t=0}^T\) satisfy \(d_t>d_{t+1}>0\). When making consumption plans, the DM initially assigns greater weight to consumption in period 1 than to consumption in any subsequent period. So, she naturally plans to consume more in period 1 than in the subsequent periods. This in turn, makes her relatively attend more to period 1. Her optimal consumption plan of period 0 should result in \(w_1/w_t>d_1/d_t\) for each \(t=2,…,T\). After arriving in period 1, we assume the DM would use the AMD factors formed in the last step as the new default discount factors. This can lead to consumption in period 1 being assigned a higher weight, and therefore create a motive for over-consumption. Such a trend could continue until she reaches the final period or runs out of money.

\begin{figure}[h] 
\vspace{12pt}
\centering 
\begin{subfigure}{0.49\textwidth} 
\centering
\includegraphics[width=\linewidth]{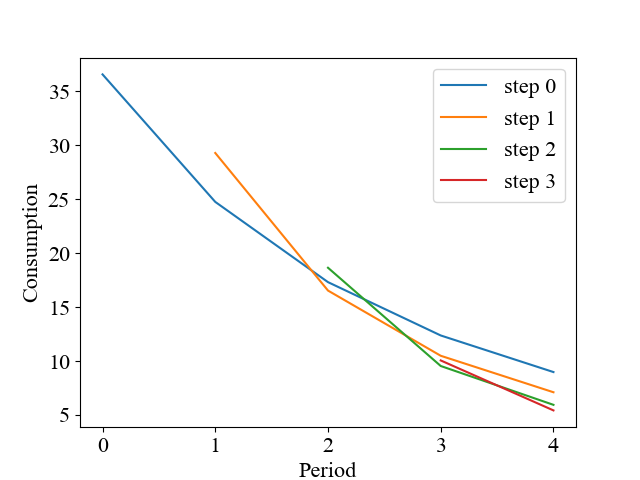} 
\caption{with learning} 
\end{subfigure} 
\hfill 
\begin{subfigure}{0.49\textwidth} 
\centering
\includegraphics[width=\linewidth]{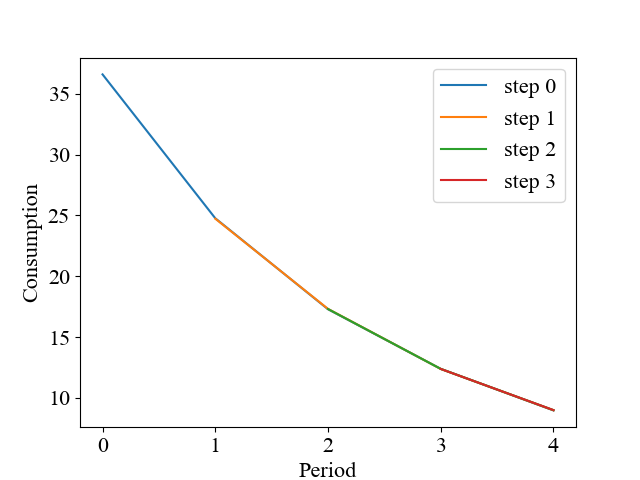} 
\caption{without learning}
\end{subfigure}
\vspace{8pt} 

\begin{minipage}{1.0\textwidth}
{\par\footnotesize Note: In step 0, the DM allocates consumption budget $m=100$ over periods 0-4. In step 1, she allocates the remaining consumption over periods 1-4, and so on. The term ``with learning'' means the DM updates default discount factors per step; ``without learning'' means the default discount factors are constant over time. $d_t^0=0.9^t$, $u(x)=x^{0.6}$, $\lambda=70$. Each optimization problem is solved by the projection gradient descent method.} 
\end{minipage}
\caption{Optimal Consumption Plan Over Time}
\label{fig:inconsistency_learning}
\end{figure}

Figure \ref{fig:inconsistency_learning} illustrates the relationship between time inconsistency and learning. Under the condition ``with learning'', which means the DM treats the AMD factors from the optimal consumption plan as default discount factors for the next step, we observe a tendency for over-consumption. Under the condition ``without learning'', which means the default discount factors are constant over time, the DM's behavior is closer to being time-consistent, and in each step, the actual consumption is slightly lower than the consumption planned earlier.\footnote{In one simulation under the condition ``without learning'', in step 0, the DM plans to consume 24.76 in period 1, and she ends up consuming 24.74. Also, in step 1, she plans to consume 17.32 in period 2, and she ends up consuming 17.31.}

We state these results formally in Proposition 8. We focus on the transition from period 0 to period 1, but the conclusions can be expanded to other periods. When the DM plans her consumption in period \(j\) (\(j=0,1\)), the default discount factor for period \(t\) is termed \(d_t^j\) and the corresponding AMD factor is termed \(w^j_t\), where \(t\geq j\). In the period 0's optimal consumption plan, the DM plans to consume \(s_t^*\) in period \(t\), and in the period 1's optimal plan, she plans to consume \(s_t^{**}\). The option space is \(A\) in period 0 and becomes \(A'\) in period 1. Notably, to perform over-consumption, the DM should not concentrate all consumption into period 0. From Proposition 7, we know that it requires \(\lambda\) to be large enough.

\noindent \textbf{Proposition 8:} \emph{Suppose the DM faces the planning problem \(\mathcal{O}(m,A,\{d^0_t\}_{t=0}^T)\) in period 0 and \(\mathcal{O}(m-s_0^*,A',\{d_t^1\}_{t=1}^T)\) in period 1. If \(\frac{d}{ds}\left(\frac{1}{u'(s)}\right)\) has a positive lower limit for \(s \in[0,m]\), there exists a threshold \(\bar{\lambda}>\underline{\lambda}\) such that when \(\lambda\in[\bar{\lambda},+\infty)\), the following is true:
\begin{enumerate} \def\labelenumi{(\alph{enumi})}
\item In period 0, for any sequence \(s_{0\rightarrow T}^*\in \{s_{0\rightarrow T}|s_{0\rightarrow T}\in A,\forall t<T:s_t>s_{t+1}>0\}\), there exist default discount factors \(\{d_t^0\}_{t=0}^T\), where \(d_t^0 >d_{t+1}^0>0\) for all \(t<T\), such that \(s_{0\rightarrow T}^*\) is the optimal consumption plan.
\item Given \(s_{0\rightarrow T}^*\) as the period 0's optimal consumption plan, if in period 1, for all \(t\geq 1\) we have \(d_t^1=d_t^0\), then there must be \(s_1^{**}< s_1^*\); if instead we have \(d_t^1=w_t^0\), then \(s_1^{**}> s_1^*\).
\end{enumerate}
}

The proof of Proposition 8 is in Online Appendix \ref{proof-8}.\footnote{As in Proposition 7, the utility-function condition in Proposition 8 is satisfied by both the logarithmic and CARA utility functions.} By statement (a), we reach an interior solution for the consumption planning problem in period 0. By statement (b), if we do not take into account the updating of default discount factors, the DM would perform under-consumption in this case. In contrast, if the default discount factors are updated based on the AMD factors formed in the most recent consumption plan, the DM would perform over-consumption. The former reflects how the DM would behave ``without learning'', while the latter reflects how the DM would behave ``with learning''.
\section{Discussion}\label{sec:model_discuss}

\subsection{Selection of Sequence Length}\label{selection-of-sequence-length}

In our model, the decision weight assigned to each period depends on the length of the sequence. As the number of periods in a sequence increases, the average decision weight for each period decreases. In Section \ref{hidden-zero-effect}, we discuss how this can generate the hidden zero effect. Nevertheless, usually in reality, we cannot observe the actual sequence length perceived by people. Researchers using our model may want to make their own assumptions about this. Here, we discuss three issues related to such assumptions and provide our recommendations.

The first issue is about the unit of time. A given duration can be represented as sequences of different lengths depending on the time unit used, such as months or days. For example, ``receive £10 in 1 month'' and ``receive £10 in 30 days'' essentially mean the same thing, but the latter seems to involve more units of time. The former can be represented by {[}0,10{]}, and the latter can be {[}0,0,\ldots,0,10{]}, with thirty zeros before the 10. In ``standard'' discounting models, the number of zeros before the 10 has no effect on the sequence value. Nonetheless, the AMD model predicts that, when the reward sequence is described with more time units, the DM may perceive the waiting period for the £10 payment as longer, thereby discounting its value to a greater extent.

As an illustration, suppose the default discount factors are exponential and the monthly discount factor is $\delta$. According to Equation (\ref{eq:w_T}), the AMD factor for the £10 payment is \(1/(1+G(T)e^{-v(10)})\). When periods are months, \(G(T)=\delta^{-1}\), whereas when periods are days, \(G(T)=\delta^{-1}+\delta^{-\frac{29}{30}}+...+\delta^{-\frac{1}{30}}\). Keeping \(\delta\) unchanged, the value of the £10 payment should be more heavily discounted in the latter case. So, if we elicit the DM's time preference from experiments and use the AMD model to fit the data, then the estimate for \(\delta\) should be greater when using days as periods than when using months as periods.

The prediction that sequences with more time units are more discounted is consistent with the numerosity effect \citep{pelham1994easy,zhang2012and,monga2012years}. This effect refers to the tendency to overestimate quantity as the number of units increases.\footnote{Some studies also suggest the unitosity effect, such as \citet{monga2012years}. For example, people may perceive ``month'' as longer than ``day''. So, when periods are months rather than days, \(\delta\) itself may be smaller.} Currently, there is no clear evidence on how the numerosity effect may affect intertemporal choices. This can be a direction for future research. In practice, we recommend that researchers choose the sequence length that matches the time unit presented to participants. For the given example, if a reward sequence is expressed as ``receive £10 in 1 month'', researchers would better represent it as {[}0,10{]} rather than a sequence with 30 zeros.


The second issue is about the end of sequence. In this paper, an implicit assumption is that each sequence terminates at the point when the final positive reward is delivered. For example, for the sequence ``receive £10 in 1 month'', we represent it as {[}0,10{]} rather than {[}0,10,0,0{]}. As is shown above, many empirical findings can be explained under this assumption. However, under this assumption, the model cannot distinguish between the values of two constant sequences of different lengths. To illustrate, consider a choice between two options: (A) receive £10 now; (B) receive £10 now, and £10 in 1 month, and £10 in 2 months. According to this assumption, we represent option A by a single-period sequence {[}10{]} and option B by {[}10,10,10{]}. In reality, people would certainly prefer option B to option A. But, as the AMD model assumes a constant sum of decision weights, both options will be equally valued at \(u(10)\).

We propose a simple remedy for this issue: researchers can assume that the DM always takes into account one additional period, or a constant number of additional periods, when valuing each sequence. For example, she represents option A as {[}10,0{]} and option B as {[}10,10,10,0{]}. Then, the value of option B would be greater than option A.\footnote{Set \(u(x)=x^{0.6}\), \(d_t=0.9^t\), \(\lambda=2\). According to the AMD model, the sequence value is 3.55 for option A and 3.84 for option B.} In Online Appendix \ref{app-adding-zero}, we show that adding a period to the end of each sequence does not affect the behavioral implications we have derived. 

\begin{figure}[h] 
\vspace{12pt}
\centering 
\includegraphics[width=0.5\textwidth]{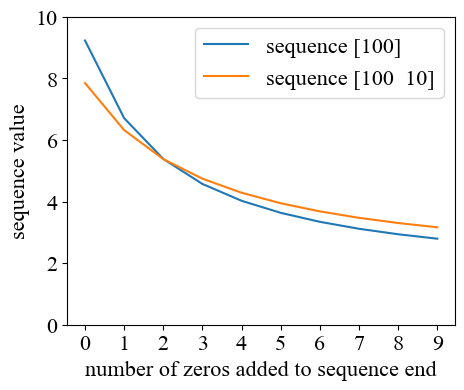} 
\vspace{8pt} 

\begin{minipage}{1.0\textwidth}
{\par\footnotesize Note: The x-axis denotes the number of additional periods added to a sequence. For example, adding two periods to the sequence [100, 10] would transform it into [100, 10, 0, 0]. $d_t^0=0.9^t$, $u(x)=x^{0.6}$, $\lambda=70$.} 
\end{minipage}
\caption{The impact of sequence length on sequence value}
\label{fig:add_zeros}
\end{figure}

The third issue is about the violation of dominance. In the AMD model, adding a tiny reward to the end of a sequence decreases the sequence value, and this is consistent with the evidence on the violation of dominance (see Section \ref{optimal-discounting}). However, this effect does not always occur. For example, when choosing between ``receive £100 now'' and ``receive £100 now and £10 in 1 month'', people would certainly prefer the latter. Such an issue can also be solved by the remedy we just proposed. Figure \ref{fig:add_zeros} illustrates how the values of these two sequences, in the AMD model, change as more zeros are added to the end of each sequence. Under the parameters in the figure, when the number of such additional zeros exceeds two, the value of the latter sequence surpasses that of the former. As the number of additional periods increases, the DM is less likely to exhibit a violation of dominance.

The rationale for adding a zero (or zeros) to the end of a sequence is as follows. When people read the phrase “receive £100 now and £10 in one month,” they not only process the information about the two payments but also infer that the sequence offers no reward beyond the one-month period, even though the latter is not explicitly described. This implies that, in evaluating the sequence, people not only pay attention to the described periods but also pay attention to information about the more distant future in order to determine when the sequence ends.

\subsection{Relation to Other Intertemporal Choice Models}\label{relation-to-other-intertemporal-choice-models}

The closest theories to our model are salience theory \citep{bordalo2012salience} and focus-weighting theory \citep{kHoszegi2013model}. Salience theory was originally proposed for risky choice but can also be applied to other decision domains \citep{bordalo2022salience}. According to salience theory, the ``salience'' of an element in a sequence increases with its deviation from a reference point. Focus-weighting theory makes a similar assumption. When the reference point is zero, both theories would predict that people pay more attention to large rewards and less attention to small rewards.

The key difference between the AMD model and these theories is that the latter do not impose a constraint on the sum of decision weights. This distinction leads to different predictions about how increasing a reward or appending a new reward affects the attention (or decision weights) allocated to other rewards in the sequence. In focus-weighting theory, increasing the attention allocated to one reward does not affect the attention allocated to other rewards, whereas in the AMD model, it reduces their attention. Salience theory incorporates this mechanism, but in that theory, appending a tiny reward to end of the sequence may increase the attention allocated to existing rewards. In the AMD model, by contrast, it necessarily reduces the attention allocated to each existing reward.

The theoretical idea that the discounting of a reward is shaped by other rewards within the same sequence originates from \citet{uzawa1968time}, a seminal paper in endogenous time preference. Early theories in this field \citep{uzawa1968time, epstein1983rate, epstein1983stationary, becker1997endogenous} assume that an increase in the present reward can lead to subsequent rewards in the same sequence being discounted more. \citet{noor2022optimal} integrate this idea with psychological mechanisms, assuming that time preferences are mediated by the cognitive cost of attending to the future. Nevertheless, these models do not specify how changes in a future reward may affect the weighting of earlier rewards (which means they cannot explain anomalies like the violation of dominance), while we fill this gap. 

Some recursive preference models \citep{epstein1989substitution, weil1990nonexpected} use a CES function to aggregate the values of current and future consumption, with the weights on the two components summing to one. This is similar to our normalization of discount factors, although we use a different framework to aggregate values.

In psychology, the intertemporal trade-off model assumes that, when valuing a sequence, people separately calculate the total utility and the utility-weighted average waiting time, and then trade off the two \citep{read2012tradeoffs,scholten2016cumulative,scholten2024unified}. This theory can also generate most of the anomalies we consider. However, it uses a lot of parameters, whereas we introduce only one parameter into the additive discounted utility framework. In addition, several studies \citep{amasino2019amount,he2023information,chen2024attribute,kang2026relationship} fit intertemporal choice data with dynamic information acquisition models, but their analysis involve only single-period rewards.



\section{Conclusive Remark}\label{sec:conclusion}

We develop a new model of time discounting based on insights into attention mechanisms. The model can be characterized by modifying the optimal discounting framework and can evolve, to a first approximation, from the learning dynamics in an SNN. We use this model to explain a range of decision anomalies.

A more important contribution of this paper, however, may be to open up new questions and stimulate further research. We note that the value of a sequence in our model is shaped by the sequence length that the
DM perceives. This raises an interesting question: how far into the future do people actually look when making decisions? Moreover, the decision weights in our model follow a softmax function, which has been widely used to model attention in various fields, such as discrete choice \citep{matvejka2015rational}, memory \citep{bordalo2020memory}, computer vision \citep{xu2015show,dosovitskiy2021image}, and language processing \citep{bahdanau2015iclr-neural,vaswani2017attention}. In machine learning, using the softmax function to model attention weights on sequence elements has become a fundamental component of large language models \citep{vaswani2017attention}. A promising direction for future research is to explore how attention mechanisms in these fields can be integrated into one unified framework.

Our model can be extended in several directions. For example, when faced with too much information, people may completely ignore some of it, so that the decision weights on some items can shrink to zero. This idea appears in the sparse max model \citep{gabaix2014sparsity}. Besides, people may allocate attention not only across periods but also across options and attributes. This can be found in theories of consideration sets \citep{manzini2014stochastic,caplin2019rational} and attribute-wise decision-making \citep{dai2014probabilistic}. Readers interested in these directions are referred to the given literature.

\clearpage
\renewcommand\refname{References}
\bibliography{reference}

\appendix
\clearpage
\begin{center}
    {\huge \textbf{Online Appendix}}
\end{center}

\section{Proof of Proposition 1}\label{proof-1}
\setcounter{equation}{0}
\renewcommand{\theequation}{A.\arabic{equation}}

We present the proof of sufficiency here. That is, if \(\succsim\) admits an optimal discounting representation and satisfies Axioms 1-4, then it admits an AMD representation.

\noindent \textbf{Lemma 1}: \emph{If Axioms 1 and 3 hold, for any \(s_{0\rightarrow T}\), there exist \(w_0, w_1, …, w_T > 0\) such that \(s_{0\rightarrow T} \sim w_0 \cdot s_0 + ...+w_T\cdot s_T\), where \(\sum_{t=0}^T w_t=1\).}

\noindent \emph{Proof}: If \(T=1\), Lemma 1 is obvious. If \(T\geq 2\), by Axiom 3, for any \(2\leq t\leq T\), there should exist \(\alpha_t\in(0,1)\) such that \(s_{0\rightarrow t}\sim \alpha_t\cdot s_{0\rightarrow t-1}+(1-\alpha_t)\cdot s_{t}\). By independence and reduction of compound alternatives, we can recursively apply such equivalence relations as follows:
\begin{equation*}
    \begin{aligned} 
    s_{0\rightarrow T} 
    &\sim \alpha_{T-1}\cdot s_{0\rightarrow T-1} + (1-\alpha_{T-1})\cdot s_T \\ 
    &\sim  \alpha_{T-1}\alpha_{T-2}\cdot s_{0\rightarrow T-2} + \alpha_{T-1}(1-\alpha_{T-2})\cdot s_{T-1} + (1-\alpha_{T-1})\cdot s_T \\ & \sim ...\\ 
    & \sim w_0 \cdot s_0 + w_1\cdot s_1 +... +w_T\cdot s_T 
    \end{aligned}
\end{equation*}
where \(w_0=\prod_{t=0}^{T-1}\alpha_t\), \(w_T = 1-\alpha_{T-1}\), and for \(0<t<T\), \(w_t=(1-\alpha_{t-1})\prod_{j=t}^{T-1}\alpha_j\). It is easy to show that the sum of \(w_0,…,w_T\) is equal to 1. \emph{QED}.

Therefore, if Axioms 1 and 3 hold, for any reward sequence \(s_{0\rightarrow T}\), we can always find a convex combination of all its elements, such that the DM is indifferent between the reward sequence and this convex combination. If \(s_{0\rightarrow T}\) is a constant sequence, i.e. all its rewards are constant, then we can directly assume \(\mathcal{W}\) is AMD-style. Henceforth, we discuss whether AMD can apply to non-constant sequences.

By Lemma 2, we show that adding a new reward to the end of \(s_{0\rightarrow T}\) has no impact on the relative decision weights in the original reward sequence.

\noindent \textbf{Lemma 2}: \emph{For any \(s_{0\rightarrow T+1}\), if \(s_{0\rightarrow T}\sim \sum_{t=0}^T w_t \cdot s_t\) and \(s_{0\rightarrow T+1} \sim \sum_{t=0}^{T+1} w'_t\cdot s_t\), where \(w_t, w'_t>0\) and \(\sum_{t=0}^Tw_t=1\), \(\sum_{t=0}^{T+1}w'_t=1\), then we can obtain \(\frac{w'_0}{w_0}=\frac{w'_1}{w_1}=…=\frac{w'_T}{w_T}\) when Axioms 1-4 hold.}

\noindent \emph{Proof}: According to Axiom 3, for any \(s_{0\rightarrow T+1}\), there exist \(\alpha,\zeta \in (0,1)\) such that
\[\label{eq:app-1-1}
\begin{aligned} s_{0 \rightarrow T}\sim\alpha\cdot s_{0 \rightarrow T-1} + (1-\alpha)\cdot s_T \\ s_{0\rightarrow T+1} \sim \zeta\cdot s_{0\rightarrow T} + (1-\zeta)\cdot s_{T+1} 
\end{aligned} 
\]
Drawing on Lemma 1, we set
\[\label{eq:app-1-2}
s_{0\rightarrow T+1} \sim \beta_0\cdot s_{0 \rightarrow T-1} + \beta_1\cdot s_T + (1-\beta_0-\beta_1)\cdot s_{T+1} 
\]
where \(\beta_0, \beta_1> 0\) and  $\beta_0+\beta_1<1$ . According to Axiom 4, \(1-\zeta=1-\beta_0-\beta_1\). So, \(\beta_0+\beta_1=\zeta\). This also implies \(\zeta > \beta_0\).

According to Axiom 2, we suppose there exists a reward sequence \(s\) such that \(s \sim \frac{\beta_0}{\zeta}\cdot s_{0 \rightarrow T-1} + (1-\frac{\beta_0}{\zeta})\cdot s_T\). By Equation \eqref{eq:app-1-2} and reduction of compound alternatives, we have \(s_{0\rightarrow T+1}\sim \zeta \cdot s + (1-\zeta)\cdot s_{T+1}\). Combining this with the second line of Equation \eqref{eq:app-1-1} and applying transitivity and independence, we obtain \(s_{0\rightarrow T} \sim \frac{\beta_0}{\zeta}\cdot s_{0 \rightarrow T-1} + (1-\frac{\beta_0}{\zeta})\cdot s_T\).

We aim to prove that for any \(s_{0\rightarrow T+1}\), we can obtain \(\alpha=\frac{\beta_0}{\zeta}\). We show this by contradiction.

Given the symmetry of \(\alpha\) and \(\frac{\beta_0}{\zeta}\), we can assume that \(\alpha > \frac{\beta_0}{\zeta}\). Consider the case that \(s_{0 \rightarrow T-1} \succ s_T\). By independence, for any reward \(c\in \mathbb{R}_{\geq 0}\), we have \((\alpha - \frac{\beta_0}{\zeta})\cdot s_{0\rightarrow T-1} + (1-\alpha+\frac{\beta_0}{\zeta})\cdot c \succ (\alpha - \frac{\beta_0}{\zeta})\cdot s_T + (1-\alpha+\frac{\beta_0}{\zeta})\cdot c\). By Axiom 2, there exists \(z\in \mathbb{R}_{\geq 0}\) such that \((1-\alpha)\cdot s_T + \frac{\beta_0}{\zeta}\cdot s_{0\rightarrow T-1}\sim z\). Given \(c\) is arbitrary, we can set \((1-\alpha+\frac{\beta_0}{\zeta})\cdot c \sim z\). By reduction of compound alternatives, we can derive that
\begin{equation*}
    (\alpha-\frac{\beta_0}{\zeta})\cdot s_{0\rightarrow T-1} +(1-\alpha)\cdot s_T + \frac{\beta_0}{\zeta}\cdot s_{0\rightarrow T-1} \succ (\alpha-\frac{\beta_0}{\zeta})\cdot s_T +(1-\alpha)\cdot s_T + \frac{\beta_0}{\zeta}\cdot s_{0\rightarrow T-1}
\end{equation*}
where the LHS can be rearranged to \(\alpha\cdot s_{0\rightarrow T-1} + (1-\alpha)\cdot s_T\), and the RHS can be rearranged to \(\frac{\beta_0}{\zeta}\cdot s_{0 \rightarrow T-1} + (1-\frac{\beta_0}{\zeta})\cdot s_1\). They should both be indifferent from \(s_{0\rightarrow T}\). This results in a contradiction.

In the case that \(s_T \succ s_{0 \rightarrow T-1}\), we can also derive such a contradiction. Meanwhile, when \(s_{0\rightarrow T}\sim s_T\), \(\alpha\) and \(\frac{\beta_0}{\zeta}\) can be any number within \((0,1)\). In that case, we can directly set \(\alpha = \frac{\beta_0}{\zeta}\).

Thus, we have \(\alpha = \frac{\beta_0}{\zeta}\) for any \(s_{0\rightarrow T+1}\), which indicates \(\frac{\beta_0}{\alpha}=\frac{\beta_1}{1-\alpha}=\zeta\). We can recursively apply this equality to any sub-sequence \(s_{0\rightarrow t}\) (\(t\leq T\)) of \(s_{0\rightarrow T+1}\), so that the lemma will be proved. \emph{QED}.

Now we move on to prove Proposition 1. The proof contains five steps.

First, we add the constraints \(\sum_{t=0}^T w_t=1\) and \(w_t>0\) to the optimal discounting problem for \(s_{0\rightarrow T}\) so that the problem can accommodate Lemma 1. According to the first-order conditions of its solution, for all \(t=0,1,….,T\), we have
\[\label{eq:app-1-3} 
f_t'(w_t)=u(s_t)+\theta 
\]
where \(\theta\) is the Lagrange multiplier. Given that \(f'_t(w_t)\) is strictly increasing, \(w_t\) is increasing with \(u(s_t)+\theta\). We define the solution as \(w_t =\phi_t(u(s_t)+\theta)\).

Second, we append a new reward \(s_{T+1}\) to end of the sequence \(s_{0\rightarrow T}\). The first-order conditions for the old periods \(t=0,1,….,T\) should remain the same form as Equation \eqref{eq:app-1-3}, but the Lagrange multiplier should change. Suppose it changes from \(\theta\) to \(\theta-\Delta \theta\). Let \(w'_0,...,w'_{T+1}\) be the solution to the optimal discounting problem for the extended sequence \(s_{0\to T+1}\), then we have \(w'_t = \phi_t(u(s_t)+\theta-\Delta \theta)\).

By Lemma 2, the relative weights of the old periods are invariant to the addition of \(s_{T+1}\). So, for any \(t=0,...,T\),
\(\frac{w_t}{w'_t}\) is equal to a constant independent of \(t\). In other words, we have
\[\label{eq:app-1-4} 
\frac{\phi_t(u(s_t)+\theta)}{\phi_t(u(s_t)+\theta-\Delta \theta)} = \kappa(\Delta \theta) 
\]
where \(\kappa(\Delta\theta)\) does not depend on \(t\). Notably, \(\kappa(\Delta\theta)\) also does not depend on the particular \(s_t\). To see this, suppose in sequence \(s_{0\to T}\), we change the reward of period \(j\) from \(s_j\) to \(s'_j\). For any other period \(t\), Equation \eqref{eq:app-1-4} and its \(\kappa(\Delta\theta)\) should not change. Given that for any \(t \neq j\), we have \(\frac{w_j}{w'_j} = \frac{w_t}{w'_t}\), this invariance should also apply to period \(j\).

Third, we show that in \(s_{0\rightarrow T}\), if we set a constant $\Delta u$ and change each \(s_t\) to \(z_t\) such that \(u(z_t)=u(s_t)+\Delta u\), the decision weights \(w_0,…,w_T\) will not be changed. Note \(\sum_{t=0}^T \phi_t(u(s_t)+\theta)=1\). It is clear that \(\sum_{t=0}^T \phi_t(u(z_t)+\theta-\Delta u)=1\). Suppose changing every \(s_t\) to \(z_t\) moves \(\theta\) to \(\theta'\) and \(\theta'<\theta-\Delta u\). Then, we must have \(\phi_t(u(z_t)+\theta')<\phi_t(u(z_t)+\theta-\Delta u)\) since \(\phi_t(.)\) is strictly increasing. Summing all such decision weights will result in \(\sum_{t=0}^T \phi_t(u(z_t)+\theta')<1\), which contradicts the constraint that the sum of decision weights is 1. The same contradiction can apply to the case that \(\theta'>\theta-\Delta u\). Therefore, changing every \(s_t\) to \(z_t\) must move \(\theta\) to \(\theta - \Delta u\), and each \(w_t\) can only be moved to \(\phi_t(u(z_t)+\theta -\Delta u)\), which is exactly the same as the original decision weight.

A natural corollary of this step is that subtracting a common number from all instantaneous utilities within a reward sequence has no effect on decision weights. For convenience, we can subtract an arbitrary number from (or add it to) the utility function.

In other words, for a given \(s_{0\rightarrow T}\) and \(s_{T+1}\), we can define a new utility function \(\mathscr{u}(.)\) such that \(\mathscr{u}(s_t) = u(s_t) +\theta-\Delta \theta\). So, \(w_t\) can be rewritten as
\[\label{eq:app-1-5} 
w_t = \frac{\phi_t(\mathscr{u}(s_t))}{\sum_{j=0}^{T}\phi_j(\mathscr{u}(s_j))} 
\]
If \(w_t\) takes the AMD form under the utility function \(\mathscr{u(.)}\), then it should also take the AMD form under the original utility function \(u(.)\).

Fourth, suppose \(\Delta \theta = \underline{\theta}\) when \(u(s_{T+1})=\underline{u}\) and \(\Delta\theta=\bar{\theta}\) when \(u(s_{T+1})=\bar{u}\), where \(\underline{u}<\bar{u}\). we show that \(\Delta \theta\) in Equation (\ref{eq:app-1-4}) has two properties: 
\begin{enumerate}
    \item[(i)] it is continuous and strictly increasing in \(u(s_{T+1})\) for \(u(s_{T+1})\in[\underline{u},\bar{u}]\);
    \item[(ii)] the width of \([\underline{\theta},\bar{\theta}]\) can exceed \(u_{\max} - u_{\min}\), where \(u_{\max}\) and \(u_{\min}\) denote the maximum and minimum instantaneous utilities in \(s_{0\to T}\).
\end{enumerate}

When these properties hold, we can manipulate \(\Delta \theta\) to any level between \([\underline{\theta},\bar{\theta}]\) by changing the newly added reward \(s_{T+1}\), with \(\bar{\theta}-\underline{\theta} \ge u_{\max} - u_{\min}\).

For property (i), we prove continuity and monotonicity separately. For monotonicity, let \(\{w'_t\}_{t=0}^{T+1}\) be the decision weights for \(s_{0\to T+1}\). Suppose \(u(s_{T+1})\) increases but \(\Delta\theta\) remains unchanged. In this case, each of \(w'_0,...,w'_T\) should also remain unchanged. However, \(w'_{T+1}\) should increase as it is strictly increasing with \(u(s_{T+1})+\theta-\Delta\theta\). This contradicts the constraint that the sum of decision weights is constant. The only way to avoid such contradictions is to set \(\Delta\theta\) strictly increasing with \(s_{T+1}\), so that \(w'_0,...,w'_T\) are decreasing with \(u(s_{T+1})\).

For continuity, define \(\mathscr{F}(\Delta\theta,u(s_{T+1})) = 
\sum_{t=0}^{T+1}\phi_t((u(s_t)+\theta-\Delta\theta)\). Then, given \(u_{T+1}\), \(\Delta \theta\) is the solution to equation \(\mathscr{F}(\Delta\theta,u(s_{T+1})) = 1\). 

Note that \(\mathscr{F}\) is continuous and increasing in \(u(s_{T+1})\), and is continuous and decreasing in \(\Delta \theta\). Moreover, \(\mathscr{F}(\underline{\theta},\underline{u}) = 1\), \(\mathscr{F}(\overline{\theta},\overline{u}) = 1\). So, for any \(u(s_{T+1})\in(\underline{u},\bar{u})\), we have \(\mathscr{F}(\underline{\theta},u(s_{T+1})) > 1\) and \(\mathscr{F}(\overline{\theta},u(s_{T+1})) < 1\). This implies that, for such \(u(s_{T+1})\), there is a unique root \(\Delta \theta \in (\underline{\theta},\overline{\theta})\) that makes \(\mathscr{F}(\Delta{\theta},u(s_{T+1})) = 1\). Therefore, for all \(u(s_{T+1})\in[\underline{u},\bar{u}]\), we define \(\Delta \theta = \mathbb{g}(u(s_{T+1}))\) and show that \(\mathbb{g}(.)\) is continuous on \([\underline{u},\bar{u}]\). 

We show this using the $\varepsilon$-$\delta$ language. Fix \(u^*\in[\underline u,\bar u]\) and let \(\theta^*=\mathbb{g}(u^*)\), so that
\(\mathscr{F}(\theta^*,u^*)=1\). Take any \(\varepsilon_{\mathrm{cont}}>0\). Since \(\mathscr{F}\) is strictly decreasing in \(\Delta\theta\), we have \(\mathscr{F}(\theta^*+\varepsilon_{\mathrm{cont}},u^*)<1\) and \(\mathscr{F}(\theta^*-\varepsilon_{\mathrm{cont}},u^*)>1\). By continuity of \(\mathscr{F}\) in both arguments, there exists \(\delta_{\mathrm{cont}}>0\) such that for all \(u^{**}\) with \(|u^{**}-u^*|<\delta_{\mathrm{cont}}\), 
\begin{equation*}
  \mathscr{F}(\theta^*+\varepsilon_{\mathrm{cont}},u^{**})<1,\quad
    \mathscr{F}(\theta^*-\varepsilon_{\mathrm{cont}},u^{**})>1. 
\end{equation*}

Take any such \(u^{**}\), and let \(\theta^{**}=\mathbb{g}(u^{**})\). By definition, \(\mathscr{F}(\theta^{**},u^{**})=1\). Since \(\mathscr{F}(\theta^*+\varepsilon_{\mathrm{cont}},u^{**})<1=G(\theta^{**},u^{**})\) and \(\mathscr{F}\) is strictly decreasing in \(\Delta\theta\), we have \(\theta^{**}<\theta^*+\varepsilon_{\mathrm{cont}}\).
Similarly, we have \(\theta^{**}>\theta^*-\varepsilon_{\mathrm{cont}}\). This suggests that \(|\mathbb{g}(u^{**})-\mathbb{g}(u^*)|=|\theta^{**}-\theta^*|<\varepsilon_{\mathrm{cont}}\).
So, \(\mathbb{g}\) is continuous at \(u^*\). Since \(u^*\in[\underline u,\bar u]\) was arbitrary, \(\mathbb{g}(.)\) is continuous on \([\underline u,\bar u]\).

A natural corollary of property (i) is that, for any \(l \in[\underline{\theta},\bar{\theta}]\), there exists \(u(s_{T+1})\in[\underline{u},\bar{u}]\) such that \(\Delta \theta = l\).

For property (ii), we show that when \(u(s_{T+1})\to \infty\), we have \(\Delta \theta \to \infty\). The proof is by contradiction. Suppose that \(\Delta\theta\) has an upper bound. Then \(u(s_t)+\theta-\Delta\theta\) is bounded below and unbounded above. 

We consider two cases. For the first case, suppose \(f'_{T+1}(.)\) has an upper bound. Then for any value of \(f'_{T+1}(w_{T+1})\), as \(u(s_{T+1})\) grows, there always exists a point at which \(u(s_{T+1})+\theta-\Delta\theta > f'_{T+1}(w_{T+1})\).
For such \(s_{T+1}\), the solution to the optimal discounting problem is \(w_{T+1}=1\) and \(w_0=\cdots=w_T=0\). However, since \(\lim_{w_t\to 0} f'_t(w_t) = -\infty\),
for \(t\le T\) and \(w_t\to 0\), we must have \(u(s_t)+\theta-\Delta\theta > f'_t(w_t)\). Thus, the derivative of the Lagrangian function with respect to \(w_t\) is positive at \(w_t=0\), so increasing \(w_t\) from zero would increase the objective. Hence \(w_t=0\) cannot be optimal, a contradiction.

For the second case, suppose \(f'_{T+1}(.)\) is unbounded above. Then as \(u_{s_{T+1}} \to 1\), we should have \(w_{T+1} \to 1\). The rest of the argument is the similar to the first case.

Finally, we derive the expression of AMD. Define \(F_t(u) = \ln \phi_t(u)\) and \(K(\Delta \theta) = \ln \kappa(\Delta \theta)\). By Equation \eqref{eq:app-1-4}, we obtain
\[\label{eq:app-1-6}
F_t(u)-F_t(u-\Delta \theta)=K(\Delta\theta)
\]
for all \(t\) and for all \(u\) in the relevant domain. Thus, for any \(\Delta \theta_1,\Delta \theta_2\), we have
\begin{equation*}
\begin{aligned}
    F_t(u)-F_t(u-\Delta \theta_1)&=K(\Delta \theta_1) \\
    F_t(u-\Delta \theta_1)-F_t(u-\Delta \theta_1-\Delta \theta_2)&=K(\Delta \theta_2)\\
    F_t(u)-F_t(u-\Delta \theta_1-\Delta \theta_2)&=K(\Delta \theta_1 + \Delta \theta_2)
\end{aligned}
\end{equation*}
By these equations, we obtain Cauchy's additive equation:
\[\label{eq:app-1-7}
K(\Delta v_1+\Delta v_2)=K(\Delta v_1)+K(\Delta v_2).
\]
Since \(f_t'(.)\) is continuous and strictly increasing, \(K(.)\) is continuous. Thus, only continuous solutions to Equation \eqref{eq:app-1-7} are linear. We can obtain \(K(\Delta \theta)=\frac{1}{\lambda} \Delta \theta\) for some constant \(\lambda>0\). 

From the fourth step, we know that \(\Delta \theta \in [\underline{\theta},\overline{\theta}]\) when \(u(s_{T+1})\in [\underline{u},\overline{u}]\), and \(\overline{\theta} -\underline{\theta}\) can exceed \(u_{\max} - u_{\min}\). Define \(\mathscr{u}_{\min} = u_{\min} + \theta - \underline{\theta}\) and set \(\mathscr{u} = \mathscr{u}_{\min} + \Delta \theta\). The linearity of \(K(.)\) yields
\[\label{eq:app-1-8}
F_t(\mathscr{u})=F_t(\mathscr{u}_{\min})+\frac{1}{\lambda}(\mathscr{u}-\mathscr{u}_{\min}).
\]
Combining Equation \eqref{eq:app-1-8} with the definition of \(F_t(.)\), we have \(
\phi_t(\mathscr{u})=\phi_t(\mathscr{u}_{\min})e^{\frac{\mathscr{u}-\mathscr{u}_{\min}}{\lambda}}\). Substituting back into Equation \eqref{eq:app-1-5} and using the corollary from the third step, we obtain
\begin{equation*}
    w_t=\frac{d_t e^{u(s_t)/\lambda}}{\sum_{j=0}^{T} d_j e^{u(s_j)/\lambda}},
\end{equation*}
with \(
d_t=\phi_t(u_{\min})e^{- u_{\min}/\lambda}\), which is exactly the AMD representation.

\section{Proof of Proposition 2}\label{proof-2}
\setcounter{equation}{0}
\renewcommand{\theequation}{B.\arabic{equation}}

Define $h_i = (x_i(\tau)-r_i)(y(\tau)-\bar{y}(\tau))$, and 
$H = \sum_{j=0}^T h_i$. By definition, from time $\tau$ to $\tau + 1$, the weight $\mathscr{w}_i$ is updated to $\mathscr{w}_i(\tau+1) = \frac{\mathscr{w}_i(\tau)+\eta h_i}{1 + \eta H}$. Since $\eta$ is small, by Taylor expansion, $\frac{1}{1+\eta H} = 1 - \eta H + \eta^2 H^2 + O(\eta^3)$. 

Let $\mathscr{W}(\tau)$ denote the set of weights at time $\tau$. At time $\tau+1$, the expected weight for signal from neuron $i$ conditional on the weights at time $\tau$, can be written as
\begin{equation}\label{eq:app-snn-1}
\begin{aligned}
 E[\mathscr{w}_i(\tau+1)|\mathscr{W}(\tau)] 
    & = E[(\mathscr{w}_i(\tau)+ \eta h_i)(1-\eta H + \eta^2 H^2 + O(\eta^3)) |\mathscr{W}(\tau)] \\
    & = \mathscr{w}_i(\tau) + \eta(E[h_i | \mathscr{W}(\tau)] - \mathscr{w}_iE[H | \mathscr{W}(\tau)]) \\
    &\qquad + \eta^2(\mathscr{w}_i E[H^2 | \mathscr{W}(\tau)] - E[h_i H | \mathscr{W}(\tau)]) + O(\eta^3)
\end{aligned}   
\end{equation}

Note that $E[H|\mathscr{W}(\tau)]=\sum_{j=0}^TE[h_j | \mathscr{W}(\tau)]$. As spike generation is independent across neurons, we have $E[h_i | \mathscr{W}(\tau)] = \mathscr{w}_i(\tau)\cdot \sigma_i^2$, where $\sigma_i^2 = r_i(1-r_i)$ is the variance of $x_i$. Also, we have 
\begin{equation*}
\begin{aligned}
       & E[h_iH|\mathscr{W}(\tau)]= \mathscr{w}_i^2 \sigma_i^2(1-3\sigma_i^2) + \sigma_i^2 \sum_{j\neq i} \mathscr{w}_j^2 \sigma_j^2 + 2\mathscr{w}_i\sigma_i^2 \sum_{j\neq i}\mathscr{w}_j\sigma_j^2 \\
       & E[H^2|\mathscr{W}(\tau)]= \sum_{i} \mathscr{w}_i^2 \sigma_i^2(1-3\sigma_i^2) + \sum_i \sigma_i^2 \sum_{j\neq i} \mathscr{w}_j^2 \sigma_j^2 + 2 \sum_{i\neq k}\mathscr{w}_i \mathscr{w}_k \sigma_i^2 \sigma_k^2
\end{aligned}
\end{equation*}
Given that $r_i$ is small and $\sigma^2_i = O(r_i)$, we can derive that $E[h_iH|\mathscr{W}(\tau)]= O(r_{\max})$ and $E[H^2|\mathscr{W}(\tau)] = O(r_{\max}(1 + \sum_{j=0}^Tr_j))$, where $r_{\max}$ is the highest firing rate among all input neurons. Equation \eqref{eq:app-snn-1} can therefore be transformed to
\[\label{eq:app-snn-2}
 E[\mathscr{w}_i(\tau+1)|\mathscr{W}(\tau)]  
 = \mathscr{w}_i(\tau) [1 + \eta (\sigma_i^2 - \bar{\sigma}^2(\tau))] + O(\eta^2 r_{\max}(1 + \sum_{j=0}^Tr_j)) 
\]
where $\bar{\sigma}^2(\tau) = \sum_{j=0}^T \mathscr{w}_j(\tau) \sigma_j^2$.

We are going to iteratively compute the expectation in Equation \eqref{eq:app-snn-2}. Before doing this, we make two approximations. First, we apply the mean-field approximation to \(E[\mathscr{w}_i(\tau) \bar{\sigma}_i(\tau) | .]\). That is, 
\[\label{eq:app-snn-3}
E[\mathscr{w}_i(\tau) \bar{\sigma}_i(\tau) | .] \approx E[\mathscr{w}_i(\tau) | .] E[\bar{\sigma}_i(\tau)|.]
\]
which introduces an error of $\text{Cov}(\mathscr{w}_i,\bar{\sigma}_i)$ to Equation \eqref{eq:app-snn-2}. 

By Taylor expansion, at each iteration, the change on $\mathscr{w}_i$ is $\Delta \mathscr{w}_i = \eta(h_i - \mathscr{w}_i H) + O(\eta^2)$. Similar to $E[h_iH|\mathscr{W}(\tau)]$ and $E[H^2|\mathscr{W}(\tau)]$, it can be shown that $\text{Var}[h_i|\mathscr{W}(\tau)]= O(r_{\max})$ and $\text{Var}[H|\mathscr{W}(\tau)]= O(r_{\max}(1 + \sum_{j=0}^Tr_j))$. Thus, we obtain $\text{Var}(\Delta \mathscr{w}_i(\tau)|\mathscr{W}(\tau)) = O(\eta^2 r_{\max}(1 + \sum_{j=0}^Tr_j))$. Furthermore, by Cauchy–Schwarz inequality, $|\text{Cov}(\mathscr{w}_i,\mathscr{w}_j)| \leq \sqrt{\text{Var}(\mathscr{w}_i)\cdot \text{Var}(\mathscr{w}_j)}$. This implies that, conditional on the weights at time $\tau - 1$, the covariance $\text{Cov}(\mathscr{w}_i,\mathscr{w}_j)$ is of the same order as $\text{Var}(\mathscr{w}_j)$ at time $\tau$. Since $\text{Cov}(\mathscr{w}_i,\bar{\sigma}_i) = \sum_{j}\text{Cov}(\mathscr{w}_i,\mathscr{w}_j) \sigma_j^2$, the order of error introduced by the mean-field approximation should be $O(\eta^2 r_{\max}(1 + \sum_{j=0}^Tr_j)(\sum_{j=0}^Tr_j))$.

Second, note that 
\begin{equation}\label{eq:app-snn-4}
\begin{aligned}
    \frac{1+\eta \sigma_i^2}{1+\eta E[\bar{\sigma}^2]} 
    & = (1+\eta \sigma_i^2)\left[1 - \eta E[\bar{\sigma}^2] + \eta^2 E[\bar{\sigma}^2]^2 + O(\eta^3)\right] \\
    & = 1 + \eta(\sigma_i^2 - E[\bar{\sigma}^2]) + \eta^2 E[\bar{\sigma}^2](E[\bar{\sigma}^2] - \sigma_i^2) + O(\eta^3) \\
    & = 1 + \eta(\sigma_i^2 - E[\bar{\sigma}^2]) + O(\eta^2 r_{\max}^2)
\end{aligned}
\end{equation}
where $E[\bar{\sigma}^2] =\sum_{j}E[\mathscr{w}_j]\sigma_j^2$.

By substituting Equations \eqref{eq:app-snn-3} and \eqref{eq:app-snn-4} into Equation \eqref{eq:app-snn-2}, we obtain
\begin{equation}\label{eq:app-snn-5}
\begin{aligned}
    E[\mathscr{w}_i(\tau)]  
    & \approx E[\mathscr{w}_i(\tau-1)]\left[1 + \eta\left(\sigma_i^2 - E[\bar{\sigma}^2(\tau-1)]\right)\right] \\
    & \approx  E[\mathscr{w}_i(\tau-1)]\frac{1+\eta \sigma_i^2}{\sum_j E[\mathscr{w}_j(\tau-1)](1+\eta \sigma_j^2)} 
    = \frac{\mathscr{w}_i(0)(1+\eta \sigma_i^2)^\tau}{\sum_j \mathscr{w}_j(0) (1+\eta \sigma_j^2)^\tau}
\end{aligned}
\end{equation}
In Equation \eqref{eq:app-snn-5}, the approximation error accumulates over iterations and becomes $O(\eta^2 \tau r_{\max}(1 + \sum_{j=0}^Tr_j)^2)$. To obtain Equation \eqref{eq:snn_weight_approx}, we need to further approximate $\sigma_i^2$ by $r_i$ and then approximate $1+\eta r_i$ by $e^{\eta r_i}$. We consider the two approximations separately.

To simplify notation, we define $\mathcal{B}^Q_i = (1+\eta \sigma_i^2)^{\tau}$, $\mathcal{B}^R_i = (1+\eta r_i)^{\tau}$, and $\mathcal{B}^S_i = e^{\eta\tau r_i}$. Accordingly, let $\mathscr{w}^Q_i = \frac{\mathscr{w}_i(0)\mathcal{B}^Q_i}{\sum_{j}\mathscr{w}_j(0)\mathcal{B}^Q_j}$, with the other terms defined similarly. We first consider the approximation $\mathscr{w}^Q_i \to \mathscr{w}^R_i$ and then consider the approximation $\mathscr{w}^R_i \to \mathscr{w}^S_i$.

For the approximation $\mathscr{w}^Q_i \to \mathscr{w}^R_i$, we rewrite $1+\eta \sigma_i^2$ as $(1+\eta r_i)(1-\epsilon^R_i)$, where $\epsilon^R_i = \frac{\eta r_i^2}{1+\eta r_i} = \eta r_i^2 + O(\eta^2 r_i^3)$. If $\tau \epsilon^R_i \ll 1$, we have 
$(1-\epsilon^R_i)^\tau = 1 - \tau \epsilon^R_i + O((\tau \epsilon^R_i)^2)$. This implies that $\mathcal{B}^Q_i \approx  \mathcal{B}^R_i(1 - \tau \epsilon_i^R)$. Thus, we obtain
\[\label{eq:app-snn-6}
\mathscr{w}^Q_i = \frac{\mathscr{w}_i(0)\mathcal{B}^Q_i}{\sum_{j}\mathscr{w}_j(0)\mathcal{B}^Q_j} 
\approx 
\frac{\mathscr{w}_i(0)\mathcal{B}^R_i(1 - \tau \epsilon^R_i)}{\sum_{j}\mathscr{w}_j(0)\mathcal{B}^R_j(1 - \tau \epsilon^R_j)}
= \mathscr{w}^R_i \cdot \frac{1 - \tau \epsilon^R_i}{1 - \tau \bar{\epsilon}^R}
\]
where $\bar{\epsilon}^R = \sum_{j} \mathscr{w}^R_j \epsilon^R_j$. If $\tau \bar{\epsilon}_i \ll 1$, by Taylor expansion, we further obtain that $\frac{1 - \tau \epsilon^R_i}{1 - \tau \bar{\epsilon}^R} \approx (1 - \tau \epsilon^R_i)(1 + \tau \bar{\epsilon}^R) \approx 1 - \tau(\epsilon^R_i - \bar{\epsilon}^R)$. Substituting the expression of $\epsilon^R_i$, the approximation $\mathscr{w}^Q_i \to \mathscr{w}^R_i$ incurs an error of order $O(\eta \tau r_{\max}^2)$.

For the approximation $\mathscr{w}^R_i \to \mathscr{w}^S_i$, note that $\ln(1 + \eta r_i) = \eta r_i - \frac{\eta^2 r_i^2}{2} + O(\eta^3 r_i^3)$. We can rewrite $(1+\eta r_i)^\tau$ as $e^{\tau \ln(1+\eta r_i)} = e^{\eta\tau r_i}\cdot e^{-\tau\epsilon^S_i}$, where $\epsilon^S_i \approx \frac{\eta^2r_i^2}{2}$. If $\tau \epsilon^S_i \ll 1$, we have $e^{-\tau\epsilon^S_i} = 1 - \tau\epsilon^S_i + O((\tau \epsilon^S_i)^2)$. This implies that $\mathcal{B}^R_i \approx  \mathcal{B}^S_i(1 - \tau \epsilon_i^S)$.
Similar to Equation \eqref{eq:app-snn-6}, we obtain $\mathscr{w}^R_i \approx \mathscr{w}^S_i \cdot \frac{1 - \tau \epsilon^S_i}{1 - \tau \bar{\epsilon}^S}$, where $\bar{\epsilon}^S = \sum_j \mathscr{w}^S_j \epsilon^S_j$. If $\tau\bar{\epsilon}^S \ll 1$, we further obtain that $\frac{1 - \tau \epsilon^S_i}{1 - \tau \bar{\epsilon}^S} \approx 1 - \tau(\epsilon^S_i - \bar{\epsilon}^S)$. Substituting the expression of $\epsilon^S_i$, the approximation $\mathscr{w}^R_i \to \mathscr{w}^S_i$ incurs an error of order $O(\eta^2 \tau r_{\max}^2)$.

Finally, we collect the conditions on $\tau$ used in the proof and conclude as follows. The expectation $E[\mathscr{w}_i(\tau)]$ admits an approximation 
\begin{equation*}
    E[\mathscr{w}_i(\tau)] \approx \frac{\mathscr{w}_i(0) e^{\eta\tau r_i}}{\sum_{j}\mathscr{w}_j(0)e^{\eta\tau r_j}}
\end{equation*}
with an error of order $O(\eta \tau r_{\max}^2 + \eta^2\tau r_{\max}(1 + \sum_{j=0}^Tr_j)^2)$. This approximation is valid provided that $\eta r_{\max}^2 + \eta^2 r_{\max}(1 + \sum_{j=0}^Tr_j)^2 \ll \frac{1}{\tau}$.

\section{Proof of Proposition 3}\label{proof-3}
\setcounter{equation}{0}
\renewcommand{\theequation}{C.\arabic{equation}}

Note the instantaneous utilities of LL and SS are \(v_l\) and \(v_s\), and the delays for LL and SS are \(t_l\) and \(t_s\). According to Equation \eqref{eq:w_T}, the common difference effect implies that, if
\[ \label{eq:app-2-1} 
\frac{v_s}{1+G(t_s)e^{-v_s}} = \frac{v_l}{1+G(t_l)e^{-v_l}} 
\]
then for any \(\Delta t \geq 0\), we have 
\[ \label{eq:app-2-2} 
\frac{v_s}{1+G(t_s+\Delta t)e^{-v_s}} < \frac{v_l}{1+G(t_l+\Delta t)e^{-v_l}} 
\]
If \(G(T)=T\), we have \(G(t+\Delta t) = G(t) + \Delta t\). In this case, combining Equation (\ref{eq:app-2-1}) and (\ref{eq:app-2-2}), we can obtain
\[ \label{eq:app-2-3}
\frac{\Delta t e^{-v_s}}{v_s} > \frac{\Delta t e^{-v_l}}{v_l} 
\]
Given that function \(\psi(v) = e^{-v}/v\) is decreasing with \(v\) so long as \(v>0\), Equation (\ref{eq:app-2-3}) is valid.

If \(G(T) = \frac{1}{1-\delta}(\delta^{-T}-1)\), we have
\[\label{eq:app-2-4} 
1+G(t+\Delta t)e^{-v} = \delta^{-\Delta t}[1+G(t)e^{-v}]+(\delta^{-\Delta t}-1)(\frac{e^{-v}}{1-\delta}-1) 
\]
Thus, combining Equation (\ref{eq:app-2-1}) and (\ref{eq:app-2-2}), we can obtain
\[\label{eq:app-2-5} 
(\delta^{-\Delta t}-1)\frac{\frac{e^{-v_s}}{1-\delta}-1}{v_s} > (\delta^{-\Delta t}-1)\frac{\frac{e^{-v_l}}{1-\delta}-1}{v_l}
\] 
Given that \(0<\delta<1\), we have \(\delta^{-\Delta t}>1\). So, Equation (\ref{eq:app-2-5}) is valid if and only if
\[\label{eq:app-2-6} 
\frac{1}{v_s}-\frac{1}{v_l}<\frac{1}{1-\delta}(\frac{e^{-v_s}}{v_s}-\frac{e^{-v_l}}{v_l}) 
\] 
By Equation (\ref{eq:app-2-1}), we know that
\[\label{eq:app-2-7} 
\frac{1}{v_s}-\frac{1}{v_l}=\frac{1}{1-\delta}\left[\frac{(\delta^{-t_l}-1)e^{-v_l}}{v_l} -\frac{(\delta^{-t_s}-1)e^{-v_s}}{v_s}\right] 
\]
Combining Equation (\ref{eq:app-2-6}) and (\ref{eq:app-2-7}), we have
\begin{equation*}
    \delta^{-t_l}\frac{e^{-v_l}}{v_l}<\delta^{-t_s}\frac{e^{-v_s}}{v_s} \Longleftrightarrow v_l - v_s + \ln \left(\frac{v_l}{v_s}\right)>-(t_l-t_s)\ln\delta
\end{equation*}

\section{Proof of Proposition 4}\label{proof-4}
\setcounter{equation}{0}
\renewcommand{\theequation}{D.\arabic{equation}}

Suppose a positive reward \(X\) is delivered at period \(T\). By Equation (\ref{eq:w_T}), if \(w_T\) is convex in \(T\), we should have \(\frac{\partial^2 w_T}{\partial T^2}\geq 0\). This implies
\[\label{eq:app-3-1} 
2G'(T)^2\geq(G(T)+e^{v(X)})G''(T) 
\]

If \(\delta=1\), then \(G(T)=T\). We have \(G'(T)=1\), \(G''(T)=0\). Thus, Equation (\ref{eq:app-3-1}) is always valid.

If \(0<\delta<1\), then \(G(T)=(1-\delta)^{-1}(\delta^{-T}-1)\). We have \(G'(T)=(1-\delta)^{-1}(-\ln\delta)\delta^{-T}\), \(G''(t)=(-\ln\delta)G'(T)\). Thus, Equation (\ref{eq:app-3-1}) is valid when 
\[\label{eq:app-3-2}
\delta^{-T}\geq(1-\delta)e^{v(X)}-1 
\]
Given \(T>0\), Equation (\ref{eq:app-3-2}) holds true in two cases. The first case is \(1\geq (1-\delta)e^{v(X)}-1\), which implies that \(v(X)\) is no greater than a certain threshold \(v(\underline{X})\), where \(v(\underline{X})=\ln(\frac{2}{1-\delta})\). The second case is that \(v(X)\) is above \(v(\underline{X})\) and \(T\) is above a threshold \(\underline{t}\). In the second case, we can take the logarithm on both sides of Equation (\ref{eq:app-3-2}). It yields \(\underline{t}=\frac{\ln[(1-\delta)\exp\{v(X)\}-1]}{\ln(1/\delta)}\).

\section{Proof of Proposition 5}\label{proof-5}
\setcounter{equation}{0}
\renewcommand{\theequation}{E.\arabic{equation}}

For convenience, we use \(v\) to represent \(v(X)\equiv u(X)/\lambda\), and use \(U\) to represent \(U(X,T)\). Set \(g= G(T)\). The first-order derivative of \(U\) with respect to \(X\) can be written as
\[\label{eq:app-4-1}
\frac{\partial U}{\partial X}=v'\frac{e^v+Ug}{e^v+g} 
\]
If \(U\) is strictly concave in \(X\), we should have \(\frac{\partial^2 U}{\partial X^2}<0\). By Equation (\ref{eq:app-4-1}), we calculate the second-order derivative of \(U\) with respect to \(X\), and rearrange this second-order condition to
\[\label{eq:app-4-2} 
2\zeta(v)+\frac{1}{1+v\zeta(v)}-1<\frac{-v''}{(v')^2}\equiv\frac{d}{dX}\left(\frac{1}{v'}\right) 
\]
where \(\zeta(v)=g/(g+e^v)\). Since \(v''<0\), the RHS of Equation (\ref{eq:app-4-2}) is clearly positive.

To prove the first part of Proposition 4, we can show that when \(X\) is large enough, the LHS of Equation (\ref{eq:app-4-2}) will be non-positive. To make the LHS non-positive, we require
\[\label{eq:app-4-3} 
\zeta(v)+\frac{1}{v}\leq\frac{1}{2} 
\]
hold true. Note that \(\zeta(v)\) is decreasing in \(v\), and \(v\) is increasing in \(X\). Hence, \(\zeta(v)+\frac{1}{v}\) is decreasing in \(X\). Besides, it approaches \(+\infty\) when \(X\rightarrow0\) and approaches 0 when \(X\rightarrow +\infty\). When \(\frac{d}{dX}\left(\frac{1}{v'(X)}\right)\) is continuous, there must be a unique realization of \(X\) in \((0,+\infty)\), say \(\bar{X}\), making the equality in Equation (\ref{eq:app-4-3}) valid. Moreover, when \(X\geq\bar{X}\), Equation (\ref{eq:app-4-3}) is always valid. In such cases, \(U(X,T)\) is concave in \(X\).

To prove the second part, first note that when \(X=0\), the LHS of Equation (\ref{eq:app-4-2}) will become \(\frac{2g}{g+1}\). If \(\frac{d}{dX}\left(\frac{1}{v'(0)}\right)\) is smaller than this number, then the LHS of Equation (\ref{eq:app-4-2}) should be greater than the RHS at the point of \(X=0\). Meanwhile, from the first part of the current proposition, we know the LHS is smaller than the RHS at the point of \(X=\bar{X}\). Thus, given \(\frac{d}{dX}\left(\frac{1}{v'(X)}\right)\) is continuous in \([0,\bar{X}]\), there must also be a point within \([0,\bar{X}]\), such that the LHS equals the RHS. Let \(X^*\) denote the minimum of \(X\) that makes the equality valid. Then, for any \(X\in(0,X^*)\), we must have that the LHS of Equation (D2) is greater than the RHS, which implies \(U(X,T)\) is convex in \(X\). Given that \(T\geq1\), we have \(g\geq1\) and thus \(\frac{2g}{g+1}\geq 1\). Therefore, when \(\frac{d}{dX}\left(\frac{1}{v'(0)}\right)<1\), \(U(X,t)\) can be convex in \(X\) for any \(X\in(0,X^{*})\), regardless of \(g\).

To prove the third part, note \(v(X)=u(X)/\lambda\). So, 
\begin{equation*}
    \frac{d}{dX}\left(\frac{1}{v'}\right)=\lambda\frac{d}{dX}\left(\frac{1}{u'}\right)
\end{equation*}
We arbitrarily draw a point from \((0,\bar{X})\) and derive the range \(\lambda\) relative to this point. For simplicity, we choose \(v=\ln g\). In this case, the LHS of Equation (\ref{eq:app-4-2}) becomes \(\frac{2}{2+\ln g}\). Define a function \(\xi(X)\), where \(\xi\) is the value of the LHS of Equation (\ref{eq:app-4-2}) minus its RHS. Note \(\xi(X)\) is continuous at \(X=\ln g\). Therefore, for any positive real number \(b\), there must exist a positive real number \(c\) such that, when \(X\in(\ln g-c,\ln g+c)\), we have
\[\label{eq:app-4-4}
    \xi(\ln g)-b<\xi(X)<\xi(\ln g)+b 
\]
If \(\xi(\ln g)-b\geq 0\), then \(\xi(X)\) will remain positive for all \(X\in(\ln g-c,\ln g+c)\), which implies the LHS of Equation (\ref{eq:app-4-2}) is always greater than its RHS.

Now we derive the condition for \(\xi(\ln g)-b\geq 0\). Suppose when \(X=\ln g\), \(\frac{d}{dX}\left(\frac{1}{u'}\right)=a\) (note at this point we have \(\frac{d}{dX}\left(\frac{1}{u'}\right)<+\infty\)). Combining with Equation (\ref{eq:app-4-3}), we know that \(\xi(\ln g)-b =\frac{2}{2+\ln g}-\lambda a-b\). Letting this value be non-negative, we obtain
\[\label{eq:app-4-5} 
\lambda \leq \frac{2}{a(2+\ln g)}-\frac{b}{a} 
\]
Given that \(T\geq1\), we have \(g\geq 1\) and thus \(\frac{2}{2+\ln g}\) should be positive. Meanwhile, given that \(u'>0\) and \(u''<0\), \(a\) should also be positive. Since \(b\) can be any positive number, Equation (\ref{eq:app-4-5}) holds if \(\lambda <\frac{2}{a(2+\ln g)}\). That is, when \(\lambda\) is positive but smaller than a certain threshold, there must be an interval \((\ln g-c,\ln g+c)\) such that the LHS of Equation (\ref{eq:app-4-2}) is greater than the RHS. Set \(X_1 = \max\{0,\ln g-c\}\), \(X_2=\min\{\bar{X}, \ln g +c\}\). When \(X\in (X_1,X_2)\), function \(U(X,T)\) must be convex in \(X\).

\section{Proof of Proposition 6}\label{proof-6}
\setcounter{equation}{0}
\renewcommand{\theequation}{F.\arabic{equation}}

The proof consists of four steps. First, we write the expressions for \(U(L1)\) and \(U(L2)\). Suppose the sequence length of each lottery is \(T\). Let \(\Omega\) denote the set of all periods with zero reward delivered, then \(\Omega=\{t|\,0\leq t\leq T,\;t \neq t_1,t_2\}\). For any \(j,k\in\{s,l\}\), we define \(\phi_j=d_{t_1}e^{v(X_j)}\) and \(\varphi_k=d_{t_2}e^{v(Y_k)}\), where \(v(s)=u(s)/\lambda\), and \(d_t\) represents the default discount factor for reward delivered at period \(t\).

For a given lottery state \((s_1,s_2)\), we denote the decision weight of each positive reward by \(w_{t_1}\) and \(w_{t_2}\). By the definition of AMD, we have
\begin{equation*}
    w_{t_1} = \frac{\phi_j}{\phi_j + \varphi_k +D} \quad ,\quad w_{t_2} = \frac{\varphi_k}{\phi_j + \varphi_k +D} 
\end{equation*} 
where \(j,k\in\{s,l\}\), \(D=\sum_{t\in\Omega} d_t\geq 0\). The value of a lottery \(L\) can be written as \(U(L)=w_{t_1}u(s_1)+w_{t_2}u(s_2)\). Hence, 
\[\label{eq:app-5-1}
\begin{aligned} 
U(L1)=0.5\frac{\phi_s u(X_s)+\varphi_s u(Y_s)}{\phi_s+\varphi_s+D} + 0.5\frac{\phi_l u(X_l)+\varphi_l u(Y_l)}{\phi_l+\varphi_l+D} \\ 
U(L2)=0.5\frac{\phi_s u(X_s)+\varphi_l u(Y_l)}{\phi_s+\varphi_l+D} + 0.5\frac{\phi_l u(X_l)+\varphi_s u(Y_s)}{\phi_l+\varphi_s+D}
\end{aligned} 
\]

Second, we observe that, when \(X_l=X_s\), we have \(U(L1)=U(L2)\). Suppose that \(X_l\) increases from \(X_s\) by a small increment. To make \(U(L1)<U(L2)\), the increment should increase \(U(L2)\) by a greater value than \(U(L1)\). Actually, we can assume \(U(L2)\) is increasing faster than \(U(L1)\) at any level of \(X_l\). That is, the partial derivative of \(U(L2)\) in terms of \(X_l\) is always greater than that of \(U(L1)\). Given \(\phi_l\) is increasing in \(X_l\), to see this, we can take partial derivatives in terms of \(\phi_l\).

In each line of Equation (\ref{eq:app-5-1}), note only the second term contains \(X_l\). Thus, we focus on the difference between the second terms. The second term of the \(U(L1)\) is influenced by \(Y_l\), while that of the \(U(L2)\) is influenced by \(Y_s\), where \(Y_l>Y_s\). Thus, we can construct a function \(\xi\) such that
\begin{equation*}
    \xi(\phi_l,\varphi) = \frac{\phi_l \cdot v(X_l)+\varphi\cdot v(Y)}{\phi_l+\varphi+D}
\end{equation*}
where \(\varphi=d_{t_2}e^{v(Y)}\). In reverse, we can define \(v(X_l)=\ln(\phi_l/d_{t_1})\) and \(v(Y)=\ln(\varphi/d_{t_2})\). The function \(\xi\) is similar to the second term of each line, but note we replace \(u(.)\) by \(v(.)\). When \(Y=Y_l\), \(\xi\) is proportional to the second term of \(U(L1)\). When \(Y=Y_s\), \(\xi\) is proportional to the second term of \(U(L2)\) (by the same proportion). Thus, to show that the partial derivative of \(U(L2)\) in terms of \(X_l\) is greater than that of \(U(L1)\), we just need to show \(\partial \xi/\partial \phi_l\) is decreasing with \(Y\) (or \(\varphi\)).

Third, we take the first- and second-order partial derivatives of \(\xi(\phi_l,\varphi)\). The partial derivative of \(\xi\) in terms of \(\phi_l\) is
\begin{equation*}
    \frac{\partial \xi}{\partial \phi_l}=\frac{(v(X_l)+1)\varphi-v(Y)\varphi+\phi_l+D(v(X_l)+1)}{(\phi_l+\varphi+D)^2} 
\end{equation*}
We need to show that for \(Y\in[Y_s,Y_l]\), we can obtain \(\partial^2 \xi/\partial \phi_l\partial \varphi<0\). This implies
\[\label{eq:app-5-2} 
(v(X_l)+v(Y)+2)D-(\phi_l-\varphi)(v(X_l)-v(Y))+2(\phi_l+\varphi)>0 
\]
We want Equation (\ref{eq:app-5-2}) to hold for any \(D\geq 0\). Given the LHS is increasing with \(D\), this can only be achieved when
\[\label{eq:app-5-3} 
2(\phi_l+\varphi)>(\phi_l-\varphi)(v(X_l)-v(Y)) 
\]
Define \(\kappa=d_{t_2}/d_{t_1}\), \(\alpha=v(X_l)-v(Y)\). Note \(\kappa\in \mathbb{R}_{>0}\), \(\alpha\in\mathbb{R}\). Equation (\ref{eq:app-5-3}) can be rewritten as
\[\label{eq:app-5-4}
(\alpha-2)\kappa^{-1} e^{\alpha}-\alpha-2<0 
\]

Fourth, based on Equation (\ref{eq:app-5-4}), we construct a function \(h(\alpha)=(\alpha-2)\kappa^{-1} e^\alpha-\alpha-2\). We aim to examine whether there exists some \(\alpha\in\mathbb{R}\) that makes \(h(\alpha)<0\). Obviously, \(\alpha=-2\) and \(\alpha=2\) satisfy this condition. Moreover, note \(h(\alpha)\) is decreasing in \(\alpha\) when \((\alpha-1)e^{\alpha}\leq \kappa\) and is increasing in \(\alpha\) otherwise. And when \(\alpha\rightarrow -\infty\) or \(\alpha \rightarrow +\infty\), we have \(h(\alpha)\rightarrow +\infty\). Thus, there must be a bounded interval \((\alpha_1,\alpha_2)\) such that \(h(\alpha)<0\) so long as \(\alpha\in(\alpha_1,\alpha_2)\), and obviously \([-2,2]\subset(\alpha_1,\alpha_2)\). Since \(v(s)=u(s)/\lambda\), we need to identify a $\lambda$ such that \(\frac{u(X_l)-u(Y)}{\lambda}\in(\alpha_1,\alpha_2)\).

For a given positive number \(\kappa\), the points \(\alpha_1,\alpha_2\) are determined by the solution to \(\frac{\alpha-2}{\alpha+2}e^{\alpha}=\kappa\). In other words, for any \(X_l\) and \(Y\in[Y_s,Y_l]\), we can always achieve \(U(L1)<U(L2)\) as long as \(u(X_l)-u(Y_l)\geq \lambda\alpha_1\) and \(u(X_l)-u(Y_s)\leq\lambda\alpha_2\). So, we can conclude that for any \(X_l>X_s>0\), \(Y_l>Y_s>0\), any sequence length of lotteries and any default discount factors (which determines \(D\) and \(\kappa\)), there exists some \(\lambda\) that makes DM intertemporal correlation averse. Specifically, all \(\lambda>\lambda^{**} ={\max}\{\frac{u(X_l)-u(Y_l)}{\alpha_1},\frac{u(X_l)-u(Y_s)}{\alpha_2}\}\) satisfy the target condition.

Notably, if \(\lambda\leq\lambda^{**}\), we have \(h(\alpha)\geq0\), which by Equation (\ref{eq:app-5-2})(\ref{eq:app-5-3}), indicates that under some conditions such as \(D=0\), there will be \(\partial^2 \xi/\partial \phi_l\partial \varphi\geq0\) for all \(Y\in[Y_s,Y_l]\). In that case, at each level of \(X_l\), the partial derivative of \(U(L1)\) in terms of \(X_l\) is greater than that of \(U(L2)\). So, increasing \(X_l\) by some increment from \(X_s\) can induce a greater increase in \(U(L1)\) than in \(U(L2)\). This makes it possible that \(U(L1)>U(L2)\). In short, DM may perform intertemporal correlation seeking when \(\lambda\leq\lambda^{**}\).

\section{Proof of Proposition 7}\label{proof-7}
\setcounter{equation}{0}
\renewcommand{\theequation}{G.\arabic{equation}}

Before proving the proposition, we first show that in the DM's optimal consumption plan \(s_{0\rightarrow T}\), the largest consumption must be \(s_0\). To show this, suppose the largest consumption is \(s_\tau\) (\(\tau>0\)). By Lemma 3 below, we can obtain that if we exchange the consumption planned in \(\tau\) with the consumption planned in period 0, the total value of consumption will be non-decreasing. So, the largest consumption must be placed in period 0.

For convenience, henceforth we use \(u_t\) to represent \(u(s_t)\).

\noindent \textbf{Lemma 3}: \emph{Suppose in} \(s_{0\rightarrow T}\)\emph{, we have} \(s_\tau= \max\{s_0,s_1,...,s_T\}\) \emph{and} \(\tau>0\). \emph{Set} \(u_0/\lambda=v_1\) \emph{and} \(u_{\tau}/\lambda=v_2\)\emph{. If we change} \(u_0/\lambda\) \emph{to} \(v_2\) \emph{and} \(u_{\tau}/\lambda\) \emph{to} \(v_1\)\emph{,} \(U(s_{0\rightarrow T})\) \emph{will be non-decreasing.}

\noindent \emph{Proof}: Let \(V/\lambda\) denote the total value of consumption before we exchange consumption between period 0 and \(\tau\), where
\begin{equation*}
    V = \frac{\sum_{t=0}^T (u_t/\lambda)\cdot d_t e^{u_t/\lambda}}{\sum_{t=0}^T d_te^{u_t/\lambda}}
\end{equation*}
Set \(d_0=\delta_1\), \(d_\tau=\delta_2\). We denote the numerator of \(V\) by \(v_1\delta_1e^{v_1}+v_2\delta_2e^{v_2}+P\) and denote its denominator by \(\delta_1e^{v_1}+\delta_2e^{v_2}+Q\).

Note \(v_2\geq V\). If changing \(u_t/\lambda\) to \(v_2\) as well as \(u_{t+\tau}/\lambda\) to \(v_1\) does not decrease \(V\), we should have 
\[\label{eq:app-6-1}
\frac{v_1\delta_1e^{v_1}+ v_2\delta_2e^{v_2}+P}{\delta_1e^{v_1}+\delta_2e^{v_2}+Q} \leq  
\frac{v_2\delta_1e^{v_2}+ v_1\delta_2e^{v_1}+P}{\delta_1e^{v_2}+\delta_2e^{v_1}+Q} 
\]
where \(\delta_1>\delta_2>0\), \(v_2\geq v_1>0\). By rearranging Equation (\ref{eq:app-6-1}), we can obtain
\[\label{eq:app-6-2} 
-(\delta_1+\delta_2)e^{v_1+v_2}(v_2-v_1)\leq e^{v_2}(Qv_2-P)-e^{v_1}(Qv_1-P) 
\]
Clearly, Equation (\ref{eq:app-6-2}) holds if \(e^v(Qv-P)\) is increasing with \(v\) when \(v\in[v_1,v_2]\), and the latter implies \(v_1\geq \frac{P}{Q}-1\).

Notably, \(V\) is a weighted mean of \(v_1\), \(v_2\) and \(\frac{P}{Q}\), and we have \(v_2\geq \max\{v_1,\frac{P}{Q}\}\). If \(v_1\geq V\), we must have \(v_1\geq \frac{P}{Q}\). In this case, Equation (\ref{eq:app-6-2}) clearly holds. If \(v_1<V\), note that \(\frac{\partial V}{\partial d_t}=\frac{\exp(u_t/\lambda)}{\sum_{j=0}^T d_j \exp(u_t/\lambda)} (u_t/\lambda-V)\). So, we will have \(\frac{\partial U}{\partial d_0}<0\) and \(\frac{\partial U}{\partial d_\tau}>0\). Both decreasing \(d_0\) to \(\delta_2\) and increasing \(d_\tau\) to \(\delta_1\) would increase the total value of consumption. In summary, for either case, after changing \(u_t/\lambda\) to \(v_2\) and \(u_{t+\tau}/\lambda\) to \(v_1\), \(V\) should be non-decreasing. \emph{QED}.

Denote the value of \(s_{0\rightarrow T}\) by \(U=\sum_{t=0}^T w_t u(s_t)\). If the optimal consumption plan is an interior solution, then the solution must satisfy \(\frac{\partial U}{\partial s_{t+1}}/\frac{\partial U}{\partial s_{t}}=1\). Suppose \(\frac{\partial U}{\partial s_t}>\frac{\partial U}{\partial s_{t+1}}>0\), then the DM can transfer a tiny consumption from \(s_{t+1}\) to \(s_t\), as increasing \(s_t\) by an increment will lead to a greater improvement in \(U\) compared to the decrease in \(U\) caused by reducing \(s_{t+1}\) by the same amount. The DM will keep transferring consumption between periods until \(\frac{\partial U}{\partial s_t}=\frac{\partial U}{\partial s_{t+1}}\). Nevertheless, if the DM keeps reducing \(s_{t+1}\) and still has \(\frac{\partial U}{\partial s_t}>\frac{\partial U}{\partial s_{t+1}}>0\) even when \(s_{t+1}\) is reduced to 0, this optimization problem will reach a corner solution. We aim to show that when \(\lambda\) is small enough, for all \(t>0\), the DM would tend to reduce \(s_t\) to zero.

The partial derivatives of \(U\) in terms of \(s_t\) is \(\frac{\partial U}{\partial s_t}=w_tu'_t(u_t+\lambda-U)\). Therefore, we have 
\[\label{eq:app-6-3}
\frac{\partial U}{\partial s_{t+1}}/\frac{\partial U}{\partial s_t} = 
\frac{d_{t+1}}{d_t} \exp\{\frac{u_{t+1}-u_t}{\lambda}\} \cdot\frac{u'_{t+1}}{u'_t} \cdot\frac{u_{t+1}+\lambda-U}{u_t+\lambda-U} 
\]
Drawing on Equation (\ref{eq:app-6-3}), we construct a function \(\rho(s_t;\mathcal{U})=e^{u_t/\lambda}u'_t(u_t+\lambda-\mathcal{U})\). Its partial derivative in terms of \(s_t\) is
\[\label{eq:app-6-4}
\frac{\partial \rho(s_t;\mathcal{U})}{\partial s_t} = 
e^{u_t/\lambda}[ (u_t+\lambda-\mathcal{U})(\frac{(u'_t)^2}{\lambda}+u''_t) +(u'_t)^2] 
\]
Observing Equation (\ref{eq:app-6-4}), we want \(\frac{(u'_t)^2}{\lambda}+u''_t \geq 0\). This inequality can be rewritten as \( \frac{1}{\lambda} \geq \frac{d}{ds_t}(\frac{1}{u_t'})\). Suppose $\frac{d}{ds}(\frac{1}{u(s)'})$ has an upper limit for \(s\in[0,m]\), we can always find a $\underline{\lambda}$ such that the inequality holds for all $\lambda < \underline{\lambda}$. 

By Lemma 3, we know that \(s_0\) must be the largest consumption in the DM's consumption plan. It can be proved that in an arbitrary sequence \(s_{0\rightarrow T}\) which satisfies this condition,  if \(\lambda\leq \underline{\lambda}\), it is always beneficial for the DM to transfer consumption from future periods to the current period, until all consumption is concentrated at the current period. We discuss this in two cases.

First, suppose for all \(t>0\), we have \(u_t+\lambda -U>0\). As \(s_0\) is the largest consumption, we must have \(u_0>U\); so, we also have \(u_0+\lambda-U>0\). In this case, if \(\lambda\leq \underline{\lambda}\), according to Equation (\ref{eq:app-6-4}), we can obtain \(\frac{\partial \rho(s_t;U)}{\partial s_t}>0\). By Equation (\ref{eq:app-6-3}), for all \(t>0\), we have \(\frac{\partial U}{\partial s_t}/\frac{\partial U}{\partial s_0}=\frac{d_{t}}{d_0} \frac{\rho(s_t;U)}{\rho(s_0;U)}\). Since \(d_0>d_t\) and \(s_0>s_t\), we can obtain \(\frac{\partial U}{\partial s_0}>\frac{\partial U}{\partial s_t}>0\). Therefore, it is beneficial for the DM to transfer a tiny consumption from \(s_t\) to \(s_0\).

Second, suppose for some period \(\tau>0\), we have \(u_\tau+\lambda-U< 0\). For this period \(\tau\), there must be \(\frac{\partial U}{\partial s_\tau}<0\). A reduction in \(s_\tau\) will increase \(U\). So, it is also beneficial to reduce it and transfer the consumption to the current period.

In conclusion, in any sequence, as long as \(s_0\) is the largest consumption, transferring consumption from the future to the present must improve \(U\). If, in such a sequence, a future period has a positive consumption, the DM will keep transferring it to the current period until she concentrates all consumption into the current period.

\section{Proof of Proposition 8}\label{proof-8}
\setcounter{equation}{0}
\renewcommand{\theequation}{H.\arabic{equation}}

We use the same notation as in the proof of Proposition 7. Before proving, we list a set of conditions that, taken together, are sufficient to derive the behavior in the proposition. We then prove that all these conditions are satisfied as long as \(\lambda\) is large enough.

First, suppose the consumption planning problem has an interior solution. Again, let \(U\) denote the value of a consumption sequence. Then, for any period \(t\) in the optimal plan, we should have \(\frac{\partial U}{\partial s_{t+1}}/\frac{\partial U}{\partial s_{t}}=1\) and \(\frac{\partial^2 U}{\partial s_t^2}<0\). The second-order condition implies
\[\label{eq:app-7-1}
(1-2w_t)\frac{1}{\lambda} + \frac{1}{u_t+\lambda-U} < -\frac{u''_t}{(u'_t)^2} 
\]

Second, to analyze how a change in \(s_t\) will affect \(\frac{\partial U}{\partial s_t}\), we construct a function \(\tilde{\rho}_t(s_0,s_1,...,s_T)\equiv\rho(s_t;U)\). By calculating the partial derivative of \(\tilde{\rho}_t\) in terms of \(s_t\), when \(u_t+\lambda-U>0\), we can obtain
\[\label{eq:app-7-2}
\frac{\partial \tilde{\rho}_t}{\partial s_t} <0 \;\Longleftrightarrow\; 
(1-w_t)\frac{1}{\lambda}+\frac{1}{u_t+\lambda-U}<-\frac{u''_t}{(u'_t)^2} 
\]

Third, when \(u_t+\lambda-U>0\), according to Equation (\ref{eq:app-6-4}), we can also obtain that \(\rho(s_t;U)\) decreases in \(s_t\) if and only if
\[\label{eq:app-7-3} 
\frac{1}{\lambda}+\frac{1}{u_t+\lambda-U}<-\frac{u''_t}{(u'_t)^2} 
\]
Clearly, if Equation (\ref{eq:app-7-3}) holds, both Equation (\ref{eq:app-7-1}) and (\ref{eq:app-7-2}) will also hold. Note \(u(m)>U\); so, if \(\lambda - u(m)\geq 0\), we will always have \(u_t+\lambda-U>0\). Under this interval of \(\lambda\), the LHS of Equation (\ref{eq:app-7-3}) is continuous and decreasing in \(\lambda\), and it converges to 0 as \(\lambda \rightarrow \infty\). Since \(-\frac{u''_t}{(u'_t)^2}\) has a positive lower bound, there must be some \(\bar{\lambda}_1 \geq u(m)\) such that, when \(\lambda\geq \bar{\lambda}_1\), for any given consumption sequence, Equation (\ref{eq:app-7-3}) is valid.

Fourth, we construct a new function \(\Gamma(s_t;\mathcal{U})=e^{u_t/\lambda}\cdot\left(\frac{u_t+\lambda}{u_t+\lambda -\mathcal{U}}\right)\). By calculating the partial derivative of \(\Gamma\) in terms of \(s_t\), when \(u_t+\lambda-U>0\), we can obtain
\[\label{eq:app-7-4} 
\frac{\partial\Gamma(s_t;U)}{\partial s_t}>0 \Longleftrightarrow (\lambda +u_t-U)^2+U(u_t-U)>0 
\]
If \(\lambda\) is large enough, Equation (\ref{eq:app-7-4}) must hold. Again, we can conclude that there is some \(\bar{\lambda}_2\geq u(m)\), such that when \(\lambda \geq \bar{\lambda}_2\), for any given sequence, Equation (\ref{eq:app-7-4}) is valid.

We define \(\bar{\lambda}\) as \(\max\{\bar{\lambda}_1,\bar{\lambda}_2\}\). It can be shown that any \(\lambda\in[\bar{\lambda},+\infty)\) satisfies the proposition.

For statement (a) in Proposition 8, note by Equation (\ref{eq:app-6-3}), we have \(\frac{\partial U}{\partial s_{t+1}}/\frac{\partial U}{\partial s_t}=\frac{d_{t+1}}{d_t} \frac{\rho(s_{t+1};U)}{\rho(s_t;U)}\). Take an arbitrary sequence from the set \(\{s_{0\rightarrow T}|s_{0\rightarrow T}\in A, \forall t<T:s_t>s_{t+1}>0\}\) (the sequence is decreasing and available for choice, and all rewards are positive). When \(\lambda \geq \bar{\lambda}\), for each period \(t<T\), we have \(\rho(s_{t+1};U)>\rho(s_t;U)\), as \(\rho(s_t;U)\) is decreasing in \(s_t\).

Denote the given sequence by \(s_{0\rightarrow T}^*\). To make \(s_{0\rightarrow T}^*\) become the interior solution for the planning problem in period 0, we need \(\frac{d_{t+1}^0}{d_t^0}=\frac{\rho(s_t;U)}{\rho(s_{t+1};U)}\). Thus, we can define \(d_t^0\) as
\begin{equation*}
    d_t^0 = \frac{\rho(s_t;U)^{-1}}{\rho(s_0;U)^{-1}+...+\rho(s_T;U)^{-1}} 
\end{equation*}
Under this definition, it can be easily validated that \(\frac{d_{t+1}^0}{d_t^0}=\frac {\rho(s_t;U)}{\rho(s_{t+1};U)}\) and \(d_t^0>d_{t+1}^0>0\). Using such \(\{d_t^0\}_{t=0}^T\), the given sequence \(s_{0\rightarrow T}^*\) is the optimal consumption plan.

For statement (b), we prove it in two steps. First, we show that if \(d_t^1=d_t^0\), the DM will under-consume in period 1. In this case, for any \(t\geq 1\) we set \(d_t^0=d_t^1=d_t\). Suppose \(s_{0\rightarrow T}^*\) is the period 0's optimal consumption plan. When moving to period 1, the DM will need to reallocate consumption over periods \(t=1,…,T\), and the largest consumption \(s_0^*\) will be moved out from the sequence. Keeping everything equal, this will reduce the sequence value \(U\), thereby (under the condition \(\lambda\geq \bar{\lambda}\)) increasing \(\frac{u_{t+1}+\lambda-U}{u_t+\lambda-U}\). Thus, according to Equation \eqref{eq:app-6-3}, \(\frac{\partial U}{\partial s_{t+1}}/\frac{\partial U}{\partial s_{t}}\) will increase to greater than 1. To optimize the consumption plan, the DM needs to adjust \(s_t\) and \(s_{t+1}\) toward re-achieving \(\frac{\partial U}{\partial s_{t+1}}/\frac{\partial U}{\partial s_{t}}=\frac{d_{t+1}}{d_t}\frac{\tilde{\rho}_{t+1}}{\tilde{\rho}_t}=1\). Specifically, she is required to reduce \(\frac{\tilde{\rho}_{t+1}}{\tilde{\rho}_t}\). According to Equation (\ref{eq:app-7-2}), when \(\lambda\geq\bar{\lambda}\), \(\tilde{\rho}_t\) is decreasing in \(s_t\), which implies that the DM needs to increase \(s_{t+1}\) in relative to \(s_t\). In other words, she needs to transfer consumption from an earlier period to a later period. So, for the optimal consumption plan in period 1, we have \(s_1^{**}<s_1^*\).

Finally, we discuss the case that \(d_t^1=w_t^0\). When moving to period 1, how the DM will adjust consumption is affected by two mechanisms. On the one hand, as \(\frac{d_{t+1}^1}{d_t^1}=\frac{d_{t+1}^0}{d_t^0}\exp\{\frac{u_{t+1}-u_t}{\lambda}\}\), the DM will initially pay more attention to an earlier period than to a later period; on the other hand, as the largest consumption \(s_0^*\) is moved out, keeping everything equal, the sequence value \(U\) will decrease. We suppose it will decrease from \(U_0\) to \(U_1\). The former mechanism drives the DM to transfer consumption from a later period to an earlier period, while the latter mechanism drives her to do the opposite. We can show that the former mechanism overrides the latter.

To show this, note for the period 0's optimal consumption plan, we have \(\frac{\partial U}{\partial s_{t+1}}/\frac{\partial U}{\partial s_{t}}=\frac{d_{t+1}^0}{d_t^0}\frac{\rho(s_{t+1}^*;U_0)}{\rho(s_t^*;U_0)}=1\). Substituting into the equation of \(\frac{\partial U}{\partial s_{t+1}}/\frac{\partial U}{\partial s_{t}}\) in period 1 (keeping everything equal), we have
\[ \label{eq:app-7-5} 
\frac{\partial U}{\partial s_{t+1}}/\frac{\partial U}{\partial s_{t}} =
\exp\{\frac{u(s_{t+1}^*)-u(s_t^*)}{\lambda}\}\frac{u_{t+1}+\lambda-U_1}{u_t+\lambda-U_1} \frac{u_t+\lambda-U_0}{u_{t+1}+\lambda-U_0}<
\frac{\Gamma(s_{t+1}^*;U_0)}{\Gamma(s_{t}^*;U_0)} 
\]
Equation (\ref{eq:app-7-5}) suggests that, under this situation, if we reduce \(U_1\) to 0, the value of \(\frac{\partial U}{\partial s_{t+1}}/\frac{\partial U}{\partial s_{t}}\) will be increased to \(\Gamma(s_{t+1}^*;U_0)/\Gamma(s_t^*;U_0)\). Since (under the condition \(\lambda \geq \bar{\lambda}\)) \(s_t^*>s_{t+1}^*\) and \(\Gamma(s_t;U_0)\) is increasing in \(s_t\), we can obtain that \(\Gamma(s_{t+1}^*;U_0)/\Gamma(s_t^*;U_0)<1\). As a result, to optimize the consumption plan in period 1, the DM will need to increase \(\frac{\partial U}{\partial s_{t+1}}/\frac{\partial U}{\partial s_{t}}=\frac{d_{t+1}^1}{d_t^1}\frac{\tilde{\rho}_{t+1}}{\tilde{\rho}_t}\). Specifically, she is required to increase \(\frac{\tilde{\rho}_{t+1}}{\tilde{\rho}_t}\). Thus, she has to transfer consumption from a later period to a later period; that is, reducing \(s_{t+1}\) in relative to \(s_t\). For period 1, we will have \(s_1^{**}>s_1^*\).

\section{Impact of Adding a Zero to the Sequence End}\label{app-adding-zero}
\setcounter{equation}{0}
\renewcommand{\theequation}{I.\arabic{equation}}

We discuss the impact of adding a period of zero reward to the end of each sequence here. This technical trick primarily affects the contexts involving choices between sequences of different lengths. For intertemporal correlation aversion, S-shaped value function, and the anomalies related to allocating resources over a certain time (e.g.~concentration bias, present bias), it doesn't require special consideration. For example, in our proof about intertemporal correlation aversion, we import a parameter \(D\) to capture the influence of all periods with zero reward delivered, and \(D\) can be arbitrary. Changing the scale of \(D\) has no effect on our proposition. We thus focus on the following anomalies: (1) common difference effect; (2) concavity of discount function.

Consider a sequence \(s_{0\rightarrow T+1}=[0,0,...,0,s_T,0]\), where \(s_T>0\) and all the other periods have no reward delivered. The discount function for \(s_T\) will be \(w_T=\frac{1}{1+G(T)e^{-v_T}}\), where \(v_T=u(s_T)/\lambda\) and 
\[\label{eq:app-8-1}
G(T)=\left\{ 
\begin{aligned} 
& \frac{\delta^{-T}-1}{1-\delta}+\delta \;,&\; 0<\delta<1 \\ 
& T+1 \;,&\; \delta=1 
\end{aligned} \right. 
\]

For common difference effect, we follow the same notation and the process of proof in Online Appendix \ref{proof-3}. If \(G(T)=T+1\), Equation (\ref{eq:app-2-3}) is still valid. Thus, the common difference effect must hold. If \(G(T)=\frac{1}{1-\delta}(\delta^{-T}-1)+\delta\), we will have
\[\label{eq:app-8-2} 
1+G(t+\Delta t)e^{-v}= \delta^{-\Delta t}(1+G(t)e^{-v})+(\delta^{-\Delta t}-1)[(\frac{1}{1-\delta}-\delta)e^{-v}-1] 
\]
which is similar to Equation (\ref{eq:app-2-4}). Combining it with Equations (\ref{eq:app-2-1})(\ref{eq:app-2-2}), we can obtain that the common difference effect holds when and only when
\begin{equation*}
    v_l-v_s+\ln\left(\frac{v_l}{v_s}\right)> \ln\left(\frac{\delta^{-t_l}-\delta(1-\delta)}{\delta^{-t_s}-\delta(1-\delta)}\right)
\end{equation*} 
 That is to say, to observe the common difference effect, the absolute and relative differences in reward utility should be significantly larger than the difference in reward delay (the implication of Proposition 3).

For concavity of discount function, we follow Online Appendix \ref{proof-4}. Again, if \(\delta=1\), Equation (\ref{eq:app-3-1}) must be valid. If \(0<\delta<1\), Equation (\ref{eq:app-3-1}) is valid when
\[\label{eq:app-8-3}
\delta^{-T}\geq (1-\delta)e^{v(x)}-1+\delta(1-\delta) 
\]
To make Equation (\ref{eq:app-8-3}) hold true, we consider two cases. The first case is that the RHS is no greater than 1. This implies that \(v(x)\leq v(\underline{x})\), where \(v(\underline{x})=\ln(\frac{2}{1-\delta}-\delta)\). If the first case does not hold, we should consider the second case: \(T\) is above a threshold \(\underline{t}\), where \(\underline{t}=\frac{\ln((1-\delta)e^{v(x)}-1+\delta(1-\delta))}{\ln(1/\delta)}\). So, when the reward is large enough (greater than \(\underline{x}\)), the discount function can be concave for short delays, as implied by Proposition 4.

\end{document}